\documentclass[brazil,12pt,a4paper,openany]{book}
\usepackage{anysize}
\usepackage[utf8]{inputenc}
\usepackage[T1]{fontenc}
\usepackage[main=english,brazil]{babel}
\usepackage[ruled,chapter]{algorithm}
\usepackage{float}
\usepackage{slashed}
\usepackage{makeidx}
\usepackage{amsmath}
\usepackage{amssymb}
\usepackage{amsthm}
\usepackage{mathrsfs}
\usepackage{graphicx}
\usepackage{caption}
\usepackage{subcaption}
\usepackage{fancyhdr}
\usepackage{array}
\usepackage{geometry}
\usepackage{simplewick}
\usepackage{latexsym}
\usepackage[all]{xy}
\usepackage{euscript}
\usepackage{enumerate}
\usepackage{dsfont}
\usepackage{slashed}
\usepackage[plainpages=false,pdftex]{hyperref}
\usepackage{pdfpages}
\usepackage{blindtext}
\usepackage{comment}
\usepackage{indentfirst,bbold}
\usepackage{physics}
\usepackage[sortcites,backref=true, sorting=none]{biblatex}

\usepackage{csquotes}

\DeclareUnicodeCharacter{0301}{\'{e}}

\makeindex

\marginsize{25mm}{25mm}{25mm}{25mm}

\hypersetup{colorlinks=true,breaklinks=true,linkcolor=black,citecolor=black,
urlcolor=black}

\begin{document}

\setlength{\abovecaptionskip}{0.0cm}
\setlength{\belowcaptionskip}{0.0cm}
\setlength{\baselineskip}{24pt}

\pagestyle{fancy}
\lhead{}
\chead{}
\rhead{\thepage}
\lfoot{}
\cfoot{}
\rfoot{}

\fancypagestyle{plain}
{
	\fancyhf{}
	\lhead{}
	\chead{}
	\rhead{\thepage}
	\lfoot{}
	\cfoot{}
	\rfoot{}
}

\renewcommand{\headrulewidth}{0pt}


\frontmatter 

\thispagestyle{empty}

\thispagestyle{empty}

\begin{figure}[h]
\center
	\includegraphics[scale=0.7]{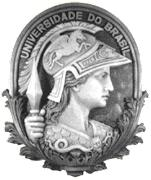}
\end{figure}

\vspace{15pt}

\begin{center}

\textbf{UNIVERSIDADE FEDERAL DO RIO DE JANEIRO}

\textbf{INSTITUTO DE FÍSICA}

\vspace{30pt}

{\Large \bf $f(R)$ Gravitation: Equivalence of Frames Upon a Conformal Transformation}

\vspace{25pt}

{\large \bf João Pedro da Cruz Bravo Ferreira}

\vspace{35pt}

\begin{flushright}
\parbox{10.3cm}{Dissertação de mestrado apresentada ao Programa de Pós-Graduação em Física do Instituto de Física da Universidade Federal do Rio de Janeiro - UFRJ, como parte dos requisitos necessários à obtenção do título de Mestre em Ciências (Física).

\vspace{18pt}

{\large \bf Orientador: Sérgio Eduardo de Carvalho Eyer Jorás}

\vspace{12pt}

{\large \bf Coorientador: Ribamar Rondon de Rezende dos Reis}}
\end{flushright}

\vspace{50pt}

\textbf{Rio de Janeiro}

\textbf{Março de 2024}

\end{center}

\newpage

\thispagestyle{empty}

\noindent

\clearpage

\newpage

\noindent

\vspace*{20pt}
\begin{center}
{\LARGE\bf Resumo}\\
\vspace{15pt}
{\Large\bf $f(R)$ Gravitation: Equivalence of Frames Upon a Conformal Transformation}\\
\vspace{6pt}
{\bf João Pedro da Cruz Bravo Ferreira}\\
\vspace{12pt}
{\bf Orientador: Sérgio Eduardo de Carvalho Eyer Jorás}\\
{\bf Coorientador: Ribamar Rondon de Rezende dos Reis}\\
\vspace{20pt}
\parbox{14cm}{Resumo da Tese de Mestrado apresentada ao Programa de Pós-Graduação em Física do Instituto de Física da Universidade Federal do Rio de Janeiro - UFRJ, como parte dos requisitos necessários à obtenção do título de Mestre em Ciências (Física).}
\end{center}
\vspace*{20pt}
Investigamos o comportamento do escalar de Ricci nas molduras de Jordan (MJ) e Einstein (ME), no contexto da gravitação $f(R)$. Discutimos a equivalência física dessas duas representações da teoria, que são matematicamente equivalentes e cujas métricas são conectadas por uma transformação conforme. Descobrimos que é possível que essa quantidade seja singular na MJ mas finita na ME, se a transformação conforme que conecta as molduras for singular no mesmo ponto que o escalar de Ricci da MJ. A ausência dessa singularidade física na ME pode ser usada como um argumento contra a equivalência física das molduras. Um gráfico do potencial na ME como função do campo conforme associado mostra que a ausência da singularidade permite que o campo assuma valores associados a valores arbitrariamente grandes do escalar de Ricci da MJ. Então, uma conjectura é proposta: a dinâmica do campo conforme pode ser interpretada como um mecanismo que pode impedir a criação de singularidades na MJ.

\vspace{4pt}

\textbf{Palavras-chave:} Gravitação $f(R)$, transformações conformes, equivalência de molduras, singularidades.



\newpage

\noindent

\vspace*{20pt}
\begin{center}
{\LARGE\bf Abstract}\\
\vspace{15pt}
{\Large\bf $f(R)$ Gravitation: Equivalence of Frames Upon a Conformal Transformation}\\
\vspace{6pt}
{\bf João Pedro da Cruz Bravo Ferreira}\\
\vspace{12pt}
{\bf Orientador: Sérgio Eduardo de Carvalho Eyer Jorás}\\
{\bf Coorientador: Ribamar Rondon de Rezende dos Reis}\\
\vspace{20pt}
\parbox{14cm}{ \emph{Abstract} da Tese de Mestrado apresentada ao Programa de Pós-Graduação em Física do Instituto de Física da Universidade Federal do Rio de Janeiro - UFRJ, como parte dos requisitos necessários à obtenção do título de Mestre em Ciências (Física).} 
\end{center}
\vspace*{20pt}
We investigate the behavior of the Ricci scalar in the Jordan (JF) and Einstein (EF) frames, in the context of $f(R)$ gravitation. We discuss the physical equivalence of these two representations of the theory, which are mathematically equivalent and whose metrics are connected by a conformal transformation. We find that it is possible for this quantity to be singular in the JF but finite in the EF, if the conformal transformation that connects frames is singular at the same point as the JF Ricci scalar. The absence of this physical singularity in the EF could be used as an argument against the physical equivalence of the frames. A plot of the EF potential as a function of the associated conformal field shows that the absence of the singularity allows the field to assume values associated to arbitrarily large values of the Ricci curvature. A conjecture is then proposed: the dynamics of the conformal field can be interpreted as a mechanism that can prevent the creation of singularities in the JF.
\vspace{4pt}




\newpage

\noindent

\vspace*{20pt}

\begin{center}

{\LARGE\bf Acknowledgements}

\end{center}

\vspace*{40pt}

I would like to thank everyone who somehow made my journey through academic life easier until this point. To all the beautiful, loving and inspiring people I have had the pleasure to cross paths with and to call friends, to call loves. Even if we are separated by space and time, I will carry each one of you in my heart. Never to be forgotten. You all know who you are. Thank you. To my mother and father, who have supported me in this beautiful, but difficult path I chose. Thank you. To my advisors, Sergio Jorás and Ribamar Reis, who have always been very understanding and have had the patience to explain the same thing to me over and over. Thank you for the knowledge you have shared with me. I also want to thank CAPES for the financial support that was crucial for me to be able to pursue this master's course until the end.


\newpage

\newpage
\phantomsection
\addcontentsline{toc}{chapter}{Contents}
\tableofcontents

\newpage
\phantomsection
\addcontentsline{toc}{chapter}{Notations and Conventions}
\section*{Notations and Conventions}
\begin{enumerate}
    \item We use natural units, in which the speed of light, the reduced Planck's constant and the gravitational constant are set to one: $c = \hbar = G = 1$.
    \item Gravitational constant $\kappa = 8\pi G = 8\pi$.
    \item Metric tensor: $g_{\mu\nu}$.
    \item Metric signature: $(-, +, +, +)$.
    \item Greek indices range from 0 to 3.
    \item Roman indices range from 1 to 3.
    \item Coordinates: $x^\mu = \{x^0, x^1, x^2, x^3\}$.
    \item Partial derivative: $\frac{\partial f}{\partial x^\mu} = \partial_\mu f = f_{,\mu}$.
    \item Time derivative: $\dot{f} = \frac{\partial f}{\partial t}$.
    \item Covariant derivative: $\nabla_\mu G^\nu = G^\nu_{;\mu} = G^\nu_{,\mu} + \Gamma^\nu_{\mu\lambda} G^\lambda - \Gamma^\lambda_{\mu\lambda} G^\nu$.
    \item Christoffel symbols: $\Gamma^\rho_{\mu\nu} = g^{\rho\sigma}(g_{\mu\sigma,\nu} + g_{\nu\sigma,\mu} - g_{\mu\nu,\sigma})$.
    \item Riemann tensor: $R^\rho_{\ \mu\kappa\nu} = \partial_\kappa\Gamma^\rho_{\mu\nu} - \partial_\nu\Gamma^\rho_{\mu\kappa} + \Gamma^\rho_{\sigma\kappa} \Gamma^\sigma_{\mu\nu} - \Gamma^\rho_{\sigma\nu} \Gamma^\sigma_{\mu\kappa}$.
    \item Ricci tensor: $R_{\mu\nu} = R^\lambda_{\ \mu\lambda\nu}$.
    \item Ricci scalar: $R = g^{\mu\nu}R_{\mu\nu}$.
    \item Energy-momentum tensor: $T_{\mu\nu}$.
    \item The function $f(R)$ derived with respect to $R$: $f'(R) =f' = \frac{df(R)}{dR}$.
    \item Second derivative of the function $f(R)$ with respect to $R$: $f''(R) = f'' = \frac{d^2f(R)}{dR^2}$.
    \item Third derivative of the function $f(R)$ with respect to $R$: $f'''(R) = f''' = \frac{d^3f(R)}{dR^3}$.
    \item In the Jordan frame: $\phi \equiv f'(R)$
    \item The tilde over a quantity (as in $\tilde{\phi}$) indicates that it is evaluated in the Einstein frame.

\end{enumerate}

\newpage
\phantomsection
\addcontentsline{toc}{chapter}{List of Figures}
\listoffigures

\mainmatter
\begin{chapter}{Introduction}
\label{intro}
$f(R)$ theories \cite{sotiriou2010f} are the simplest alternative to Einstein's general relativity \cite{einstein1916foundation} proposed to explain inflation and/or cosmic acceleration as a purely gravitational effect, without the need for an exotic component. The fundamental equations of the theory are obtained by a simple modification of the Einstein-Hilbert action, in which the Ricci scalar ($R$) is replaced by a function $f(R)$ of said scalar \cite{sotiriou2010f}. While Einstein's General Relativity (GR) provides us with a description of Nature that explains very well effects such as gravitational lensing, gravitational time dilation and many aspects of cosmology, it does not inherently explain early time cosmic inflation and the current cosmic acceleration.

The introduction of an exotic fluid or a cosmological constant is a way around this gap in the theory that allows predictions in accordance with observations, but even then the theory lacks a fundamental explanation for the mechanism that generates these aspects of Nature. An alternative gravitational theory must then be able to explain all that GR does and also aspects that it does not. Among the many possible ways to modify gravity, the introduction of an $f(R)$ is the simplest one. The requirements for a new theory mentioned above imply that it must generate a cosmological evolution in accordance with what is observed. This means that the generated universe needs to go through the epochs of radiation and matter domination and lead the way for the cosmic acceleration we see today to emerge naturally.

The recovery of the correct cosmology is no guarantee that the $f(R)$ is valid, that is because a new degree of freedom emerges in this modified gravity as a consequence of the relation between the Ricci scalar and the trace of the energy-momentum tensor, $T$, now being differential and not purely algebraic, as in GR. More explicitly, a scalar field $\phi$ is defined as the derivative of the $f(R)$ with respect to $R$ and it obeys a differential equation in which the behavior of the field is dependent on $T$ and will, in general, modify the gravitational force between bodies. This new aspect of the force can generate, for instance, significant deviations on the trajectories of bodies in the scale of our solar system, leading to inconsistencies between the movements predicted by the theory and movements of the planets that are very well described by GR. Fortunately, these deviations can be suppressed by what is known as the chameleon effect \cite{khoury2004chameleon}, a mechanism through which the range of the field changes as a function of the energy-matter density present in that region of space. This mechanism provides strong constraints on parameters of the theory \cite{pretel2020strongest}.

$f(R)$ theories can be expressed in two representations that are connected by a conformal transformation (CT) on the metric. They are called Jordan's (JF) and Einstein's frame (EF) and, while being mathematically equivalent, the physicality of the Einstein frame is an ongoing matter of discussion because of an anomalous coupling between the conformal field and matter. Such coupling modifies the equation for conservation of energy and momentum, causing the movement of test particles to deviate from the geodesics of the EF metric. In this dissertation we aim to explore the issue of the physical equivalence between the frames by studying the behavior of the Ricci scalar and its singularities across them. Since this is a scalar quantity, the associated singularities are physical, being coordinate independent and representing a true aspect of reality where physics, as we know it, breaks down. The main question we seek to answer is: is it possible for a scalar quantity like $R$ to present a singularity in one frame but not in the other? If so, can the frames be considered physically equivalent?

This dissertation is structured as follows. In Chapter \ref{chapter2} we introduce General Relativity and $f(R)$ theories. The equations of the theory are obtained using the Lagrangian formalism and a crucial difference between the theories, associated to the new scalar degree of freedom in $f(R)$ is pointed out. After this, the minimum conditions that must be satisfied by the derivatives of $f(R)$ for the theory to cosmologically viable will be discussed.

In Chapter \ref{chapter3} a discussion on coordinate transformations and singularities is held. We begin with a discussion on how the importance of coordinate systems changes across Newtonian Physics, Special Relativity and General Relativity. After this, we discuss the difference between physical and coordinate singularities, showing how the latter can be removed from the metric by a convenient coordinate transformation that is singular at the same point as the original metric, leading to a line element in which only the physical singularity remains. This discussion is important because a parallel will be drawn in Chapter \ref{chapter5}, when we consider the behavior of the physical singularity of $R$ upon a CT.

Chapter \ref{chapter4} covers the subject of conformal transformations in the context of $f(R)$ theories. It begins with a mathematical introduction, focusing on how CTs preserve angles, and thus, the causal structure of spacetime and then follows to explain how CTs are different from coordinates transformations. Next, a classic application of CTs in electrostatics is discussed, emphasizing how the preservation of angles has powerful consequences, allowing for a problem to be formulated in a more convenient way. After this, we describe how the scalar degree of freedom $f(R)$ can be explored in order to cast the theory either in the JF or the EF. The field equations in both frames are obtained and the apparent equivalence between $f(R)$ and Brans-Dicke theory is discussed. After this, a discussion on how the units in the EF being a function of the spacetime point can be used as an argument for the physical equivalence of the frames is held \cite{faraoni2007pseudo}.

In Chapter \ref{chapter5}, we explore how a physical singularity of the JF Ricci scalar does not manifest in its EF counterpart if the conformal transformation connecting the frames is singular at the same point as the JF scalar. This analysis, conducted for two $f(R)$ models, one exponential and one Starobinsky-like, suggests a conjecture: the dynamics of the conformal field, when interpreted through the perspective of the JF, could be related to the creation of singularities during stellar collapse. The investigation of this conjecture is left as a possibility for future work.

Finally, in Chapter \ref{conclusions}, we give final remarks and present our conclusions.
\end{chapter}
\begin{chapter}{Field Equations Of Gravitation}
\label{chapter2}

\hspace{5 mm}

$f(R)$ theories provide us with more general field equations than Einstein's that reduce themselves to the standard equations from GR when $f(R) = R$. Here we calculate Einstein's equation and then proceed to obtain the generalized ones using the Lagrangian formalism.
\section{Einstein's Field Equations}\label{sec:einstein_eq}
Einstein's equations can be obtained by varying the Einstein-Hilbert action with respect to the inverse metric $g^{\mu\nu}$ and making use of the principle of least action. The action is given by
\begin{equation}
    S = \frac{1}{2\kappa}\int{d^4x \, \sqrt{-g}R} + S_m,
    \label{eq:hilbert_action}
\end{equation}
where $g$ is the determinant of the metric $g_{\mu\nu}$, $S_m$ the matter action and $R = g^{\mu\nu} R_{\mu\nu}$ the Ricci scalar. Taking the variation of eq. (\ref{eq:hilbert_action}) with respect to $g^{\mu\nu}$ yields
 \begin{subequations} 
\begin{equation}
    \delta S = (\delta S)_1 + (\delta S)_2 + (\delta S)_3 + \delta S_m,
    \label{subeq:deltaS}
\end{equation}
\begin{equation}
 (\delta S)_1 \equiv \int{d^4 x\, \sqrt{-g}\,g^{\mu\nu}\delta R_{\mu\nu}},
 \label{subeq:deltaS1}
\end{equation}
\begin{equation}
(\delta S)_2 \equiv \int d^4 x\,\sqrt{-g}\,R_{\mu\nu}\,\delta g^{\mu\nu},
\label{subeq:deltaS2}
\end{equation}
\begin{equation}
(\delta S)_3 \equiv \int d^4 x\,R\,\delta (\sqrt{-g}). 
\label{subeq:deltaS3}
\end{equation}
\end{subequations}
First we will consider $(\delta S)_1$, $(\delta S)_2$ and $(\delta S)_3$, and leave the matter term for last. In order to make use of the least action principle we need to write each term in terms of $\delta g^{\mu\nu}$. $(\delta S)_2$ is already in this form, so we need to develop $(\delta S)_1$ and $(\delta S)_3$. We begin with the former.

First we explicitly write the Riemann tensor in terms of a small variation with respect to the affine connection. This gives us
    \begin{align}
        R_{\mu \lambda \nu}^{\rho} + \delta R_{\mu \lambda \nu}^{\rho} = \; &\partial_{\lambda}(\Gamma_{\nu \mu}^{\rho} + \delta\Gamma_{\nu \mu}^{\rho}) + (\Gamma_{\lambda\sigma}^{\rho} + \delta\Gamma_{\lambda\sigma}^{\rho})(\Gamma_{\nu\mu}^{\sigma} + \delta\Gamma_{\nu\mu}^{\sigma}) \notag \\
        &- \partial_{\nu}(\Gamma_{\lambda \mu}^{\rho} + \delta\Gamma_{\lambda \mu}^{\rho}) - (\Gamma_{\nu\sigma}^{\rho} + \delta\Gamma_{\nu\sigma}^{\rho})(\Gamma_{\lambda\mu}^{\sigma} + \delta\Gamma_{\lambda\mu}^{\sigma}).
        \label{eq:riemannvariation}
    \end{align}
    Now we make use of a locally inertial frame, this means that at some point $p$ the metric is equivalent to Minkowski's and its first derivatives at that point are zero. Remember that tensorial equations obtained with this argument will be valid in other reference frames, since tensorial equations are independent of coordinates \cite{carroll2019spacetime}. From this argument and the expression for the connection $\Gamma_{\alpha\beta}^{\gamma}$, we have that $\partial_{\mu} g_{\alpha\beta} = 0 \Rightarrow \Gamma_{\alpha\beta}^{\gamma} = 0$. Notice that, while $\Gamma_{\alpha\beta}^{\gamma} = 0$, the same is not true for its derivatives, $\partial_{\nu}\Gamma_{\alpha\beta}^{\mu}$, since they are related to second derivatives of the metric that do not have to be zero at locally inertial frames, unless spacetime is flat, which is not our case. Expression (\ref{eq:riemannvariation}) then becomes
\begin{align}
      R_{\mu \lambda \nu}^{\rho} + \delta R_{\mu \lambda \nu}^{\rho} &= \partial_{\lambda} \Gamma_{\nu\mu}^{\rho} + \partial_{\lambda} \delta\Gamma_{\nu\mu}^{\rho} + \delta\Gamma_{\lambda\sigma}^{\rho} \delta\Gamma_{\nu\mu}^{\sigma} - \partial_{\nu} \Gamma_{\lambda\mu}^{\rho} - \partial_{\nu} \delta\Gamma_{\lambda\mu}^{\rho} - \delta\Gamma_{\nu\sigma}^{\rho} \delta\Gamma_{\lambda\mu}^{\sigma} \notag \\
      &= \left[\partial_{\lambda} \Gamma_{\nu\mu}^{\rho} - \partial_{\nu} \Gamma_{\lambda\mu}^{\rho}\right] + \left[\partial_{\lambda} \delta\Gamma_{\nu\mu}^{\rho} - \partial_{\nu} \delta\Gamma_{\lambda\mu}^{\rho}\right] + \mathcal{O}(\delta^2),
      \label{eq:rieman_variation_second_order}
\end{align}
in which $\mathcal{O}(\delta^2)$ contains the terms proportional to the square of the variation. Since $\Gamma_{\alpha\beta}^{\gamma} = 0$ in a locally inertial frame, the covariant derivatives reduce to partial ones, $\nabla_{\alpha} = \partial_{\alpha}$. As can be seen in equation (\ref{subeq:deltaS1}), we are interested in the variation of $R_{\mu\nu}$, so we need to contract $\rho$ with $\lambda$ in equation (\ref{eq:rieman_variation_second_order}). Discarding the second order terms, the variational term becomes 
\begin{align}
       \delta R_{\mu\nu} = \delta R_{\mu \lambda \nu}^{\lambda} = \nabla_{\lambda} \delta\Gamma_{\nu\mu}^{\lambda} - \nabla_{\nu}\delta\Gamma_{\lambda\mu}^{\lambda}, 
       \label{eq:delta_Ricci_tensor}
\end{align}
notice again that since this is a tensorial equation, it is valid in any frame. We now substitute this result in equation (\ref{subeq:deltaS1}) for $(\delta S)_1$, which yields
\begin{align}
(\delta S)_1 &= \int d^4 x \sqrt{-g} g^{\mu\nu} \left[ \nabla_{\lambda} (\delta \Gamma^\lambda_{\nu\mu}) - \nabla_\nu (\delta \Gamma^\lambda_{\mu\lambda}) \right] \notag \\
&= \int d^4 x \sqrt{-g} \nabla_\sigma \left[ g^{\mu\nu} (\delta \Gamma^\sigma_{\mu\nu}) - g^{\mu\sigma} (\delta \Gamma^\lambda_{\mu\lambda}) \right].
\label{eq:deltaS1.2}
\end{align}
Now we need the variation of the connection with respect to $g^{\mu\nu}$. The expression of $ \Gamma_{\mu\nu}^{\sigma}$ in terms of $g_{\mu\nu}$ gives us
\begin{align}
        \Gamma_{\mu\nu}^{\sigma} + \delta \Gamma_{\mu\nu}^{\sigma} &= \frac{1}{2}(g^{\sigma\rho} + \delta g^{\sigma\rho})(\partial_{\mu}g_{\nu\rho} + \partial_{\mu}\delta g_{\nu\rho} + \partial_{\nu}g_{\rho\mu} + \partial_{\nu}\delta g_{\rho\mu} - \partial_{\rho}g_{\mu\nu} - \partial_{\rho}\delta g_{\mu\nu}) \notag \\ \notag
        &= \frac{1}{2}\left[ g^{\sigma\rho}(\partial_{\mu}g_{\nu\rho} + \partial_{\nu}g_{\rho\mu} - \partial_{\rho}g_{\mu\nu}  ) + g^{\sigma\rho}(\partial_{\mu}\delta g_{\nu\rho} + \partial_{\nu}\delta g_{\rho\mu} - \partial_{\rho}\delta g_{\mu\nu}) + \right. \\ 
        &\left.\delta g^{\sigma\rho}(\partial_{\mu}g_{\nu\rho} + \partial_{\nu}g_{\rho\mu} - \partial_{\rho}g_{\mu\nu}) + 
        \delta g^{\sigma\rho}(\partial_{\mu}\delta g_{\nu\rho} +  \partial_{\nu}\delta g_{\rho\mu} - \partial_{\rho}\delta g_{\mu\nu})\right],
\end{align}
again, using the locally inertial frame, we have that $\Gamma_{\alpha\beta}^{\gamma} = 0$. Discarding $\mathcal{O}(\delta^2)$ terms leaves us with 
    \begin{align}
        \delta \Gamma_{\mu\nu}^{\sigma} = \frac{1}{2}g^{\sigma\rho}(\nabla_{\mu}\delta g_{\nu\rho} + \nabla_{\nu}\delta g_{\rho\mu} - \nabla_{\rho}\delta g_{\mu\nu}).
        \label{eq:delta_Gamma}
    \end{align}
Using the relation $\delta g_{\mu\nu} = -g_{\mu\rho}g_{\nu\sigma}\delta g^{\sigma\rho}$ \cite{carroll2019spacetime}, between the variation of the metric and its inverse and remembering that the covariant derivative of the metric is zero, we obtain
    \begin{align}
      \delta \Gamma_{\mu\nu}^{\sigma} &=  \frac{1}{2}g^{\sigma\rho}[\nabla_{\mu}(-g_{\nu\alpha}g_{\rho\beta}\delta g^{\alpha\beta}) + \nabla_{\nu}(-g_{\rho\alpha}g_{\mu\beta}\delta g^{\alpha\beta}) -\nabla_{\rho}(-g_{\mu\alpha}g_{\nu\beta}\delta g^{\alpha\beta})] \notag \\
      &= \frac{1}{2}g^{\sigma\rho}(-g_{\nu\alpha}g_{\rho\beta}\nabla_{\mu}\delta g^{\alpha\beta} -g_{\rho\alpha}g_{\mu\beta}\nabla_{\nu}\delta g^{\alpha\beta} +g_{\mu\alpha}g_{\nu\beta}\nabla_{\rho}\delta g^{\alpha\beta}) \notag \\
      &= \frac{1}{2}(-g_{\nu\alpha}\delta_{\beta}^{\sigma}\nabla_{\mu}\delta g^{\alpha\beta} -g_{\mu\beta}\delta_{\alpha}^{\sigma}\nabla_{\nu}\delta g^{\alpha\beta} +g_{\mu\alpha}g_{\nu\beta}\nabla^{\sigma}\delta g^{\alpha\beta}) \notag \\
      &\Rightarrow \delta \Gamma_{\mu\nu}^{\sigma} = -\frac{1}{2}(g_{\lambda\mu}\nabla_{\nu}\delta g^{\lambda\sigma} +g_{\lambda\nu}\nabla_{\mu}\delta g^{\lambda\sigma} -g_{\mu\alpha}g_{\nu\beta}\nabla^{\sigma}\delta g^{\alpha\beta})
      \label{eq:gamma_variation}.
\end{align}
We can now substitute (\ref{eq:gamma_variation}) in (\ref{eq:deltaS1.2}). Calculating term by term:
\begin{align}
        g^{\mu\nu}\delta \Gamma_{\mu\nu}^{\sigma} &= -\frac{1}{2}g^{\mu\nu}(g_{\lambda\mu}\nabla_{\nu}\delta g^{\lambda\sigma} + g_{\lambda\nu}\nabla_{\mu}\delta g^{\lambda\sigma} - g_{\mu\alpha}g_{\nu\beta}\nabla^{\sigma}\delta g^{\alpha\beta} ) \notag 
        \\ 
        &= -\frac{1}{2}(\delta_{\lambda}^{\nu}\nabla_{\nu}\delta g^{\lambda\sigma} + g_{\lambda\nu}\nabla^{\nu}\delta g^{\lambda\sigma} - g_{\nu\beta}\delta_{\alpha}^{\nu}\nabla^{\sigma}\delta g^{\alpha\beta} ) \notag
        \\
        &= -\frac{1}{2}(\nabla_{\lambda}\delta g^{\lambda\sigma} + g_{\lambda\nu}\nabla^{\nu}\delta g^{\lambda\sigma} - g_{\alpha\beta}\nabla^{\sigma}\delta g^{\alpha\beta} ),\notag
        \\ 
        -g^{\mu\sigma}\delta \Gamma_{\lambda\mu}^{\lambda} &= \frac{1}{2}g^{\mu\sigma}(g_{\phi\lambda}\nabla_{\mu}\delta g^{\phi\lambda} + g_{\phi\mu}\nabla_{\lambda}\delta g^{\phi\lambda} - g_{\lambda\alpha}g_{\mu\beta}\nabla^{\lambda}\delta g^{\alpha\beta} ) \notag
        \\
        &= \frac{1}{2}(g_{\phi\lambda}\nabla_{\sigma}\delta g^{\phi\lambda} + \nabla_{\lambda}\delta g^{\sigma\lambda} - g_{\lambda\alpha}\nabla^{\lambda}\delta g^{\alpha\sigma})\notag 
        \\
        &= \frac{1}{2}(g_{\phi\lambda}\nabla_{\sigma}\delta g^{\phi\lambda} + \nabla_{\lambda}\delta g^{\sigma\lambda} - \nabla_{\lambda}\delta g^{\alpha\sigma} ),  \notag  
        \\
        \Rightarrow g^{\mu\nu}\delta \Gamma_{\mu\nu}^{\sigma} &- g^{\mu\sigma}\delta \Gamma_{\lambda\mu}^{\lambda} = g_{\alpha\beta}\nabla^{\sigma}\delta g^{\alpha\beta} - \nabla_{\lambda}\delta g^{\sigma\lambda}.
        \label{eq:delta_gamma_to_metric}
\end{align}
Plugging this into expression (\ref{eq:deltaS1.2}) for $(\delta S)_1$ gives us 
\begin{equation}
    (\delta S)_1 = \int{d^4 x \sqrt{-g} \nabla_{\sigma}\left[g_{\alpha\beta}\nabla^{\sigma}\delta g^{\alpha\beta} - \nabla_{\lambda}\delta g^{\sigma\lambda}\right]}.
\end{equation}
We then notice that this is the integral, in a certain volume, of the covariant divergence of a vector,  which is equivalent to a boundary contribution by the curved space Stokes' theorem, 
\begin{equation}
\int_{\Sigma} \nabla_{\nu} V^{\nu} \sqrt{-g} d^n x = \int_{\partial \Sigma} n_{\nu} V^{\nu} \sqrt{-\gamma} d^{n-1} x,
\label{eq:stokes}
\end{equation}
where $V^{\mu}$ is a vector field over a region $\Sigma$, $n_{\mu}$ is normal to the boundary $\partial \Sigma$ and $\gamma_{\mu\nu}$ is the induced metric on the boundary. Such contribution can be set to zero by making the variation vanish at infinity \footnote{Actually, the boundary term will include a variation of the metric's first derivative, which is usually not set to zero. This contribution can be cancelled by adding an extra term to the action \cite{poisson2004relativist}.}.
    
Moving on to $(\delta S)_3$, we need the expression for $\delta\sqrt{-g}$. Here we use the following result, valid for any square matrix $M$ with nonvanishing determinant: $\ln(\mathbf{det} M) = \mathbf{Tr}(\ln M)$ \cite{carroll2019spacetime}. The variation with respect to $M$ gives us
    \begin{equation}
        \frac{1}{\mathbf{det}M}\delta(\mathbf{det}M) = \mathbf{Tr}[M^{-1}\delta M].
    \end{equation}
In terms of the metric and its determinant, we have
\begin{align}
  \delta g &= g (g^{\mu\nu} \delta g_{\mu\nu}) \notag \\
        &= -g(g_{\mu\nu}\delta g^{\mu\nu}). 
\end{align}
The expression for $ \delta \sqrt{-g}$ is then
 \begin{align}
        \delta \sqrt{-g} &= \frac{1}{2\sqrt{-g}}\delta g \notag \\
       &= \frac{1}{2\sqrt{-g}}gg_{\mu\nu}\delta g^{\mu\nu} \notag \\
       &=-\frac{1}{2}\sqrt{-g}g_{\mu\nu}\delta g^{\mu\nu}.
       \label{eq:sqrt_variation}
\end{align}
Now that $(\delta S)_1$, $(\delta S)_2$ and $(\delta S)_3$ are expressed in terms of $\delta g^{\mu\nu}$, we are ready to obtain the field equations by making use of the least action principle. The functional derivative of the action satisfies
\begin{equation}
\delta S = \int{d^4x \, \frac{\delta S}{\delta g^{\mu\nu}}\delta g^{\mu\nu}},
\label{eq:functional_derivative}
\end{equation}
and the least action principle says that for stationary points $\frac{\delta S}{\delta g^{\mu\nu}} = 0$. In terms of the results obtained for $(\delta S)_1$, $(\delta S)_2$ and $(\delta S)_3$, we have
\begin{align}
    \delta S = \frac{1}{2\kappa}\int{d^4x \, \sqrt{-g}\left(R_{\mu\nu} - \frac{1}{2}R g_{\mu\nu}\right)\delta g^{\mu\nu}} + \delta S_m.
\end{align}
Comparing this with (\ref{eq:functional_derivative}) we see that
\begin{equation}
    \frac{\sqrt{-g}}{2\kappa}\left(R_{\mu\nu} - \frac{1}{2}R g_{\mu\nu}\right) + \frac{\delta S_m}{\delta g^{\mu\nu}} = 0,
\end{equation}
and defining the matter energy-momentum tensor as 
\begin{equation}
    T_{\mu\nu}^{(m)} = \frac{-2}{\sqrt{-g}}\frac{\delta S_m}{\delta g^{\mu\nu}}
    \label{eq:energy_momentum_tensor},
\end{equation}
gives us the Einstein's field equations for the metric:
\begin{equation}
R_{\mu\nu} - \frac{1}{2}R g_{\mu\nu} = \kappa T_{\mu\nu}^{(m)},
\label{eq:einstein_eq}
\end{equation}
and the trace of this equation gives us
\begin{equation}
    R = -\kappa T^{(m)},
\end{equation}
where we see that, in Einstein's theory, the relation between $R$ and $T^{(m)}$ is algebraic. This is a major difference between GR and $f(R)$ theories, where said quantities are related by a differential equation, highlighting a new scalar degree of freedom present in the theory, as will be seen in the next section.

\section{The $f(R)$ Field Equations}\label{sec:f(R)_equations}
The $f(R)$ field equations can be obtained by following the same procedure shown in section \ref{sec:einstein_eq}. We begin with the action of the theory, which is obtained by using a function $f(R)$ in place of $R$ in the Einstein-Hilbert action:
\begin{equation}
    S =\frac{1}{2k}\int{d^4x \, \sqrt{-g}f(R)} + S_m.
    \label{eq:f(R)_action}
\end{equation}
Like before, we need to vary this action with respect to $g^{\mu\nu}$, which gives us
\begin{equation}
   \delta S =  \frac{1}{2k}\int{d^4x\, [ \delta \sqrt{-g}f(R) + \sqrt{-g}\delta f(R)}] + \delta S_m,
    \label{eq:f(R)_variation1}
\end{equation}
the expression for $\delta \sqrt{-g}$ was obtained previously in (\ref{eq:sqrt_variation}). To compute the variation of the $f(R)$ we make use of the chain rule:
\begin{align}
    \delta f(R) &= \frac{df(R)}{dR}\delta R \notag \\
    &= f'(R) \delta(g^{\mu\nu}R_{\mu\nu}) \notag \\
    &= f'(R) \left(R_{\mu\nu}\delta g^{\mu\nu} + g^{\mu\nu} \delta R_{\mu\nu}\right)
    \label{eq:delta_f(R)}
\end{align}

Substituting $\delta \sqrt{-g}$ e $\delta f(R)$ in (\ref{eq:f(R)_variation1}) gives us
\begin{equation}
    \delta S = \frac{1}{2k}\int{d^4x \, \sqrt{-g}\left[-\frac{1}{2}f(R)g_{\mu\nu}\delta g^{\mu\nu} + f_R R_{\mu\nu}\delta g^{\mu\nu} + f_R g^{\mu\nu}\delta R_{\mu\nu}\right]} +\delta S_m,
    \label{eq:f(R)variation2}
\end{equation}
Notice that we still need to express the third term in terms of $\delta g^{\mu\nu}$. Equation (\ref{eq:delta_Ricci_tensor}) for $\delta R_{\mu\nu}$ in terms of the connection gives us
\begin{align}
    g^{\mu\nu}\delta R_{\mu\nu} &=  g^{\mu\nu}\left(\nabla_{\lambda} \delta\Gamma_{\mu\nu}^{\lambda} - \nabla_{\mu}\delta\Gamma_{\nu\lambda}^{\lambda}\right) \notag \\
    &= \nabla_{\lambda}(g^{\mu\nu}\delta \Gamma_{\mu\nu}^{\lambda}) - \nabla_{\mu}(g^{\mu\nu}\delta \Gamma_{\nu\lambda}^{\lambda})\notag \\
    &= \nabla_{\alpha}\left(g^{\mu\nu}\delta \Gamma_{\mu\nu}^{\alpha} - g^{\mu\alpha}\delta \Gamma_{\mu\nu}^{\nu}\right),
    \label{eq:g_delta_R}
\end{align}
with this result we can develop the third term inside the square brackets of equation (\ref{eq:f(R)variation2}), which becomes
\begin{equation}
 \int{d^4x \, \sqrt{-g}f'(R) g^{\mu\nu} \delta R_{\mu\nu}} =  \int{d^4x \, \sqrt{-g}f'(R) \nabla_{\alpha}\left(g^{\mu\nu}\delta \Gamma_{\mu\nu}^{\alpha} - g^{\mu\alpha}\delta \Gamma_{\mu\nu}^{\nu}\right)}.
\end{equation}
The integral of $\nabla_{\alpha}\left[f'(R)\left(g^{\mu\nu}\delta \Gamma_{\mu\nu}^{\alpha} - g^{\mu\alpha}\delta \Gamma_{\mu\nu}^{\nu}\right)\right]$ is, by Stoke's theorem, equivalent to a surface term which we assume to be zero at infinity, and thus, integration by parts gives us
\begin{align}
 \int{d^4x \, \sqrt{-g}f'(R) \nabla_{\alpha}\left(g^{\mu\nu}\delta \Gamma_{\mu\nu}^{\alpha} - g^{\mu\alpha}\delta \Gamma_{\mu\nu}^{\nu}\right)} &= \int{d^4x \, \sqrt{-g} \left(g^{\mu\alpha}\delta \Gamma_{\mu\nu}^{\nu} - g^{\mu\nu}\delta \Gamma_{\mu\nu}^{\alpha}\right)\nabla_{\alpha}f'(R)} \notag \\
 &= \int{d^4x \,\sqrt{-g} \left(\nabla_{\nu}\delta g^{\alpha\nu} - g_{\sigma\beta}\nabla^{\alpha}\delta g^{\sigma\beta}\right)\nabla_{\alpha}f'(R)}, \notag \\
 &= \int{d^4x \, \sqrt{-g}}\left(g_{\mu\nu}\square - \nabla_{\mu}\nabla_{\nu}\right)f'(R), \label{eq:f_R_delta_R}
\end{align}
where $\square \equiv \nabla^{\mu}\nabla_{\mu}$ is the covariant d'Alembertian. In the passage to the second line we used equation (\ref{eq:delta_gamma_to_metric}) to express variations of the connection in terms of the inverse metric and from the second to the third line we rewrote the integrand using the following relations:
\begin{align}
&\nabla^{\alpha}\left[g_{\sigma\beta} \delta g^{\sigma\beta}\nabla_{\alpha}f'\right] = \nabla^{\alpha}(g_{\sigma\beta} \delta g^{\sigma\beta}) \nabla_{\alpha}f' + g_{\sigma\beta} \delta g^{\sigma\beta}\square f' \label{eq:box_f(R)_calc1},  \\
&\nabla_{\nu}\left[\nabla_{\alpha}f' \delta g^{\alpha\nu}\right] = \nabla_{\alpha} f' \nabla_{\nu} \delta g^{\alpha\nu} + \delta g^{\alpha\nu}\nabla_{\nu}\nabla_{\alpha}f'.\label{eq:box_f(R)_calc2}
\end{align}
Notice that, when integrated, the left side of equations (\ref{eq:box_f(R)_calc1}) and (\ref{eq:box_f(R)_calc2}) will also be equivalent to surface terms by Stoke's theorem and once again we set them to zero by demanding that the variation vanishes at infinity. Substituting (\ref{eq:f_R_delta_R}) into (\ref{eq:f(R)variation2}) and evoking the least-action principle, as done in the previous section, leads us to the $f(R)$ field equations, also known as generalized Einstein's equations:
\begin{equation}
    f'(R)R_{\mu\nu} - \frac{1}{2}f(R)g_{\mu\nu} + (g_{\mu\nu}\square - \nabla_{\mu}\nabla_{\nu})f'(R) = kT_{\mu\nu}^{(m)}
    \label{eq:f(R)_equation}, 
\end{equation}
where $T_{\mu\nu}^{(m)}$ was defined in the previous section. Notice that since $R$ is written in terms of second derivatives of the metric, the term inside the parentheses on the left-hand side of (\ref{eq:f(R)_equation}) will be proportional to the fourth derivative of the metric, and thus we see that these are fourth-order partial differential equations for the metric components. Notice also that, if the action is linear in $R$, the fourth-order terms will vanish and we recover general relativity. A major difference between $f(R)$ and GR is how the presence of matter and energy affects the curvature of spacetime. Recall that in GR the trace of Einstein's equation yields $R = -kT^{(m)}$, where the Ricci curvature is algebraically determined by the trace of $T_{\mu\nu}^{(m)}$, whereas the trace of equation (\ref{eq:f(R)_equation}) is
\begin{equation}
    f'(R)R - 2f(R) + 3\square f'(R) = kT^{(m)}.
    \label{eq:f(R)_trace}
\end{equation}
We see that, in this case, $R$ is a dynamical variable related to $T$ through a differential equation where $T^{(m)}$ acts as a source and not algebraically, as in GR. This new scalar degree of freedom will be explored when we discuss conformal transformations.

\subsection{Conditions On $f'(R)$ And $f''(R)$}
When testing $f(R)$ theories, it is vital that not only they reproduce the successes of GR, but also that they do not introduce instabilities that go against experimental observations. For this reason, we will look at a procedure established in \cite{faraoni2006matter}, where an instability in the gravitational sector for the function $f(R) = R - \mu^4/R$, discovered in \cite{dolgov2003can}, was generalized for an arbitrary $f(R)$ function. This was done by demanding stability of the theory under perturbations relative to GR, resulting in conditions for $f''(R)$ that rule out certain models.

To describe perturbations around GR, the weak field limit will be used, meaning the metric is approximately flat and the Ricci scalar is written as follows
\begin{align}
    &g_{\mu\nu} = \eta_{\mu\nu} + h_{\mu\nu},
    \label{eq:metric_perturbation} \\
    &R = 0 + R_1,
    \label{eq:R_perturbation}
\end{align}
with $\eta_{\mu\nu}$ being the Minkowski metric with signature $(-1,1,1,1)$, and $h_{\mu\nu}$ and $R_1$, the perturbations with respect to $\eta_{\mu\nu}$ and its associated curvature, respectively. Notice that the GR value for $R$ in the Minkowski metric is 0, which is why it was explicitly included in (\ref{eq:R_perturbation}). This is different from what is done in reference \cite{faraoni2006matter}, where the metric is written in the same way, but it is still assumed that the unperturbed value of the curvature is $R = -\kappa T$, which is not consistent with the fact that the unperturbed metric is flat. Our $f(R)$ is written as
\begin{equation}
    f(R) = R + \Delta (R),
    \label{eq:f(R)_perturbation}
\end{equation}
in which the deviation from GR is contained in the function $\Delta(R)$. The weak field limit allows us to write the d'Alembertian $\square$ as 
\begin{equation}
    \square f' = -\ddot{f'} + \nabla^2 f',
    \label{eq:dalambertian_perturb}
\end{equation}
where $\ddot{f'} = \partial_t^2 f'$ and $\nabla^2 = \partial^{i}\partial_{i}$ is the Laplacian in three-dimensional Euclidean space. Plugging (\ref{eq:R_perturbation}) and (\ref{eq:f(R)_perturbation}) in (\ref{eq:dalambertian_perturb}) and discarding second order terms in $h_{\mu\nu}$ and $R_1$ gives us
\begin{align}
    &\ddot{f'} =  \Delta''(R)\ddot{R}_1, 
    \label{eq:dalamb_temporal} \\
    &\nabla^2 f' = \Delta''(R) \nabla^2 R_1,
    \label{eq:dalamb_spacial}
\end{align}
Substituting (\ref{eq:dalamb_temporal}) and (\ref{eq:dalamb_spacial}) in (\ref{eq:f(R)_trace}), the trace of the $f(R)$ field equations yields the following differential equation for the perturbation $R_1$,
\begin{equation}
    \ddot{R}_1 - \nabla^2R_1 + \frac{1 - \Delta'}{3\Delta''}R_1 = - \frac{2\Delta}{3\Delta''}.
    \label{eq:R1_oscilator}
\end{equation}
Notice that the coefficient of $R_1$, on the third term on the left-hand side can be interpreted as the square of an effective mass \cite{pogosian2008pattern}. Since $\Delta (R)$ is a perturbation, we have that $f'(R) = 1 + \Delta' \approx 1$, which means that $1-\Delta' > 0$, leaving the sign of the effective mass term purely dependent on $\Delta''(R)$. We see then, that the theory will be stable if $\Delta''(R) > 0$, and unstable otherwise. The effective mass term being negative is associated to instabilities because it generates exponential solutions for (\ref{eq:R1_oscilator}) that grow indefinitely. 

A condition for $f'(R)$ can also be obtained in the context of conformal transformations applied to $f(R)$. In Chapter \ref{chapter4} we will see that $f'(R)$ plays the role of a conformal factor that is applied to the metric with the objective of writing the action of the theory in a representation called the Einstein Frame (EF), where the field equations become those of GR with the presence of a conformal field. The conformal transformation is defined in such a way that the condition $f'(R) > 0$ has to be satisfied. Also, this is required to avoid anti-gravity, which is associated to a negative effective gravitational constant, as shown in \cite{amendola2010dark}.
\end{chapter}
\begin{chapter}{Coordinate Transformations}
\label{chapter3}
In order to better understand future arguments about singularities/divergences in the context of conformal transformations, it is important to first get a better understanding of said topics in the context of coordinate transformations. In this Chapter we discuss the role of coordinates in gravitational theories and the existence of coordinate and physical singularities.
\section{Coordinate Systems}
For a better understanding of the role of coordinates in General Relativity, $f(R)$ and other geometrical theories of gravity, it is important to state how this role changes in the context of Newtonian Mechanics, Special Relativity and GR. The Newtonian principle of absolute and independent space and time states that the time and space intervals between any two events are the same with respect to any frame. In this context space and time are absolute \cite{marion2013classical} in the sense that coordinates are simply used to label pre-existing points in a fixed space where physical events take place and such events could never change the fundamental structure of space. While this is true, spatial coordinates are relative in the sense that the positions of particles depend on the reference frame used. Another important aspect is the existence of inertial reference frames, in which laws are always expressed in the same form \cite{taylor2005classical}. In this context, the only possible coordinate transformations from one reference frame to another, that preserve the form of the equations of motion, are simple shifts in positions and velocities, plus spatial rotations. Also, the time coordinate in the new frame can never depend on the spatial coordinates of the old frame \cite{fukushima1986coordinate}, these are known as Galilean transformations.

In the case of Special Relativity (SR), is it postulated that the speed of light must be the same in all inertial frames. This forces one to abandon the notion that time and space are absolute and independent from each other, giving rise to the notion of a unified 4-dimensional entity called spacetime. An immediate consequence is that the notion of simultaneity is no longer absolute and observers at different speeds will measure different temporal and spatial separation between events on spacetime. The notion of inertial frames still holds, but in this case the allowed coordinate transformations, called Lorentz transformations \cite{schutz2022first}, are more general than those of Newtonian Physics and can mix temporal and spatial coordinates.

In GR, the Einstein Equivalence Principle states that the laws of physics must reduce to those of special relativity in small enough regions of space, and that it is impossible to detect gravitational fields by means of local experiments \cite{carroll2019spacetime}. This implies that the speed of light must be invariant under any changes of the 
reference frames, not restricted to the inertial ones, which in turn implies that the allowed coordinate transformations to switch between reference frames are more general than those of SR, with any smooth coordinate transformation being allowed. Another crucial difference between GR and SR is that in the former, the geometry of spacetime depends on mass and energy (because of this, the notion of inertial frames is now valid only locally), which generate curvature, which is perceived as gravity. There is no correct set of coordinates, but the study of different phenomena may benefit from different coordinate systems and care must be taken when interpreting certain results. Since the curvature of spacetime depends on the physical phenomena taking place, there are extreme cases in which this curvature is infinite, these are called physical singularities, and signal a drawback in the theory. Not every singularity is physical, they can occur purely because our coordinates fail to correctly describe the physics at some point. A specific example that illustrates well how coordinates can be misleading is the case of Schwarzschild black holes, which will be discussed in the next section.
An interesting discussion on importance of coordinates in the context of Newtonian physics, SR and GR can be found in reference \cite{fukushima1986coordinate}.
\section{The Schwarzschild Solution}
One of the most important solutions for Einstein's equations is the Schwarzschild metric. It is an unique exterior vacuum solution \cite{carroll2019spacetime} for the case of a spherically symmetric and static distribution of mass $M$. It has the following line element in spherical coordinates:
\begin{align}
    &ds^2 = -\left(1 - \frac{2M}{r}\right) dt^2 + \left(1 - \frac{2M}{r}\right)^{-1} dr^2 + r^2 d\Omega^2, \label{eq:schwar_metric} \\
    &d\Omega^2 = d\theta^2 + \sin^2{\theta}d\phi^2. \notag
\end{align}
$M$ is identified as the mass an observer at infinity would measure. The quantity $R_S = 2M$ is called the Schwarzschild radius and plays an important role when we are describing a black hole in this geometry, as will be seen.
\section{Singularities}
Notice that equation (\ref{eq:schwar_metric}) is singular at $r = 0$ and $r = 2M$, meaning they either become null or diverge to infinity. When $r = 0$ we see that $g_{tt}$ diverges and $g_{rr}$ is null. When $r = 2M$ the opposite happens. The nature of these singularities is not clear at first: one cannot tell if they are physical or just a result of the chosen coordinates failing to describe spacetime at these points. How does one distinguish between physical and coordinate singularities? The answer to this lies in scalars, quantities whose value does not depend on the coordinate system used. More specifically, curvature scalars can be used to check for physical singularities. It is worth noting that physical singularities can be defined in a way more general than just the divergence of scalars. They can also be defined in terms of geodesics that have a finite affine length. Such geodesics are called incomplete or inextendible,  since they cannot exist at a point of the manifold where a singularity is present 
\cite{wald2010general}. However, in this work, we will refer to singularities simply as the divergence of curvature scalars. The Ricci scalar is one of the most fundamental scalars that can be constructed from the metric and its first and second derivatives. It directly reflects how the curvature of spacetime is being affected by any matter distribution. In our case, its value is $R = 0$ because the Schwarzschild metric is a vacuum solution and the trace of Einstein's equation, (\ref{eq:einstein_eq}), demands that this scalar vanishes. This, however, does not mean that there is no curvature in spacetime, as can be seen by considering the contraction of the Riemann tensor with itself, known as the Kretschmann scalar. We have:
\begin{equation}
 K = R^{\mu\nu\alpha\beta}R_{\mu\nu\alpha\beta} = \frac{48M^2}{r^6}.
 \label{eq:schwar_kret_scalar}
\end{equation}
We see that (\ref{eq:schwar_kret_scalar}) diverges at $r = 0$, indicating the location of the only physical singularity, so the singularity of (\ref{eq:schwar_metric}), at the Schwarzschild Radius $R_S$ may be a coordinate one, meaning that there must be a different set of coordinates that are well behaved at that point. Before we move on to investigating such coordinates, let us stick to the spherical ones and ask ourselves if there is anything special happening at $r = R_S$. To get a better understanding of what goes on at that location, we can look at the causal structure associated to light cones. We need to consider radial null curves with $\theta$ and $\phi$ constant and $ds^2 = 0$. This gives us
\begin{equation}
    \frac{dt}{dr} = \pm \left(1 - \frac{2M}{r}\right)^{-1},
    \label{eq:spherical_light_cones}
\end{equation}
which tells us that, in a $t-r$ diagram, for $r \rightarrow \infty$, the slope of light cones is $\pm 1$ and as $r \rightarrow 2M$, it goes to $\pm \infty$, meaning that they close up as we approach $R_S = 2M$. This is shown in figure \ref{fig:tr_diagram_spherical}, which indicates that a test particle would take infinite coordinate time to reach $R_S$ because the closing up of light cones can be wrongly interpreted to mean that no particle would ever cross the line $r = 2M$.
\setlength{\abovecaptionskip}{10pt}
\begin{figure}[H]
    \centering
    \includegraphics[scale=0.5]{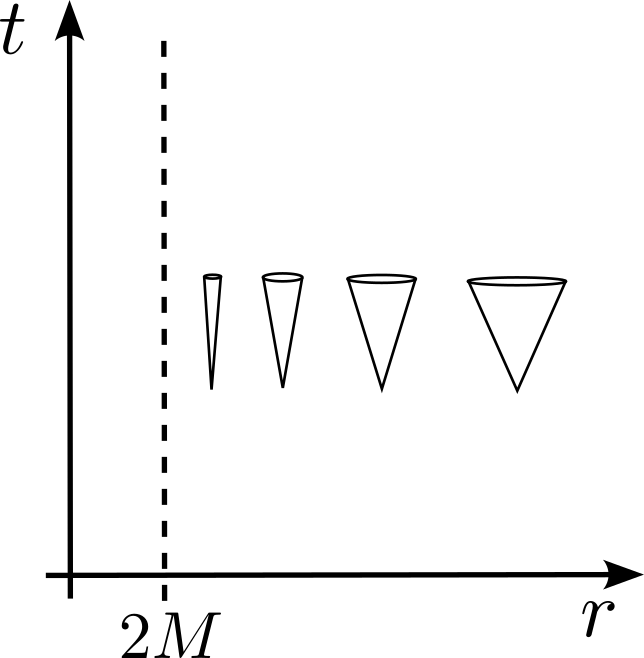}
    \caption{The closing up of light cones in spherical coordinates shows that no particle would ever cross the surface $r = 2M$ in a finite coordinate time.}
    \label{fig:tr_diagram_spherical}
\end{figure}
We have already seen through equation (\ref{eq:schwar_kret_scalar}) that there is a physical singularity at $r = 0$ and that there should be nothing wrong at $R_S$. To see how this $t-r$ diagram can be misleading let's look at the proper time it would take for a particle of mass $m$ coming from rest at infinity to fall radially to the surface $r = 2M$. Reference \cite{schutz2022first} shows how the symmetries of the Schwarzschild metric allows us to write the following equations for the components of the 4-momentum:
\begin{align}
p^{t} &= m \left(1 - \frac{2M}{r}\right)^{-1}, \\
p^{r} &= m \frac{dr}{d\tau}, \\
p^{\phi} &= 0, \\
p^{\theta} &= 0, 
\end{align}
this along with the fact that $\Vec{p}\cdot \Vec{p} = -m^2$ leads to:
\begin{equation}
    d\tau = -\frac{dr}{\left(\frac{2M}{r}\right)^{1/2}}.
\end{equation}
Integrating both sides from $2M$ to $r = R \gg 2M$ gives us
\begin{equation}
    \Delta \tau = \frac{4M}{3} \left[ \left(\frac{r}{2M}\right)^{3/2} \right]_{2M}^{R},
    \label{eq:proper_time_infall}
\end{equation}
which clearly is finite, confirming that the diagram in figure \ref{fig:tr_diagram_spherical} misleads one into thinking the particle would never reach the surface $r = 2M$. In fact, we can even change the lower limit in \ref{eq:proper_time_infall} to some radius $r < 2M$ or even $r = 0$, showing not only that the particle would reach $R_S$ in a finite proper time, but also that it would cross this surface and reach the singularity at $r = 0$. We have seen that the Schwarzschild metric has a coordinate singularity at $R_S$ but that, physically, there is nothing wrong with spacetime there, one should then be able to find coordinates that are well suited to describe the same physics and present no coordinate singularity at $R_S$.
\section{The Kruskal-Szekeres Coordinates}
The coordinates we seek were proposed independently in 1960 by Kruskal \cite{kruskal1960maximal} and Szekeres \cite{szekeres1960singularities} and are called the Kruskal-Szekeres coordinates. They describe spacetime for $r > 2M$, $r < 2M$ and $r = 2M$. From this point on we will refer to the surface located at $R_S$ as the event horizon, the reason for this will be clear shortly. The new coordinates are the timelike $T$ and the spacelike $R$ and are given by 
\begin{align}
T &= \left( \frac{r}{2M} - 1 \right)^{1/2} e^{r/4M} \sinh\left( \frac{t}{4M} \right), \label{eq:kruskal_coords_T} \\
R &= \left( \frac{r}{2M} - 1 \right)^{1/2} e^{r/4M} \cosh\left( \frac{t}{4M} \right),
\label{eq:kruskal_coords_R}
\end{align}
for $r > 2M$, and
\begin{align}
T &= \left(1 - \frac{r}{2M}\right)^{1/2} e^{r/4M} \cosh\left(\frac{t}{4M}\right), \\
R &= \left(1 - \frac{r}{2M}\right)^{1/2} e^{r/4M} \sinh\left(\frac{t}{4M}\right),
\end{align}
for $r < 2M$. Now $r$ is now a function of $T$ and $R$ given by
\begin{equation}
    T^2 - R^2 = \left(1 - \frac{r}{2M}\right) e^{r/2M}
    \label{eq:kruskal_r}
\end{equation}
Notice that these transformations are singular at $r = 2M$, but this is necessary to remove the coordinate singularity there. This becomes clear when we express $ds$ in the new coordinates:
\begin{equation}
    ds^2 = \frac{32M^3}{r} e^{-r/2M} (-dT^2 + dR^2) + r^2 d\Omega^2.
    \label{eq:kruskal_metric}
\end{equation}
Here it is clear that the only singularity present is the physical one, at $r = 0$ and so a $T-R$ diagram must give a better picture of what actually happens at $r = 2M$ than the one in figure \ref{fig:tr_diagram_spherical}. In fact, the radial null curves which define the slope of light cones are
\begin{equation}
    \frac{dT}{dR} = \pm 1,
    \label{eq:kruskal_light_slope}
\end{equation}
which shows clearly how their slope are now constant. Also, from (\ref{eq:kruskal_r}) we see that surfaces of constant $r$ will appear as hyperbolas in the $T-R$ diagram and dividing (\ref{eq:kruskal_coords_T}) by (\ref{eq:kruskal_coords_R}) we see that those of constant $t$ are given by
\begin{equation}
    \frac{T}{R} = \tanh{\left(\frac{t}{4M}\right)},
\end{equation}
which will appear as straight lines passing through the origin. Notice that for $t \rightarrow \infty$ we have $T = R$, and the same happens for $r = 2M$ in (\ref{eq:kruskal_r}). This shows, in accordance to what we have seen in the last section, that $r = 2M$ and $t = \pm \infty$ define the surface at $R_S$. The $T-R$ diagram is shown in figure \ref{fig:kruskal_diagram}. In it we can see the hyperbolas corresponding to constant $r$ and straight lines of constant $t$. The dashed line represents a radially infalling particle, and as we have seen from (\ref{eq:kruskal_light_slope}), every null ray is at $45^{\circ}$. It's important to keep in mind that every point in the diagram is a 2-sphere of events, since only $R$ and $T$ were plotted and the angular coordinates $\theta$ and $\phi$ were suppressed. The null lines passing through the origin divide spacetime in 4 regions. Region I corresponds to $r > 2M$, where the Schwarzschild coordinates are well defined. Notice that the physical singularity in $r = 0$ is the future of every light cone inside region II, which is considered to be the "inside" of a black hole. The line $r = 2M$, $t = +\infty$ is what separates the rest of spacetime from this point of no return, from which anything that enters is directed to the singularity, this line, which is actually a null surface is what we call an event horizon, and a black hole is anything that is separated from spacetime by an event horizon. Covering regions III and IV is out of the scope of this work but more detail about them can be found in \cite{carroll2019spacetime}.
\begin{figure}[H]
    \centering
    \includegraphics[scale=0.5]{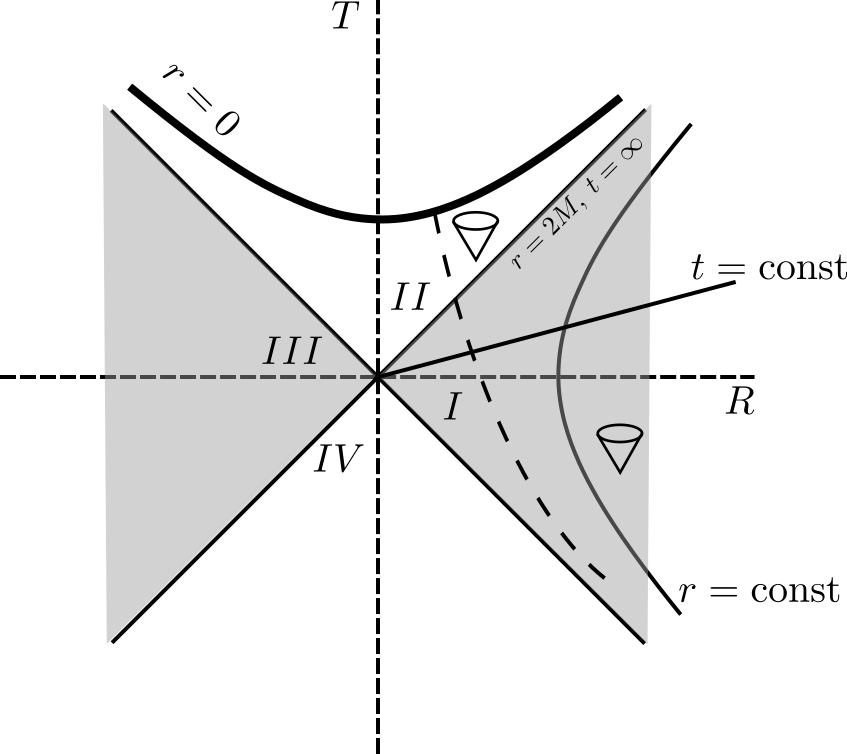}
    \caption{$T-R$ diagram. The hyperbolas are curves of constant $r$ and the straight lines, of constant t. The line $r = 2M$, $t = \infty$ correspond to the event horizon, a null surface which when crossed inevitably leads any particle or light to the physical singularity at $r = 0$.}
    \label{fig:kruskal_diagram}
\end{figure}
We have seen how different sets of coordinates can be used to describe the same physical situation, but that care must be taken when interpreting results, since the presence of coordinate singularities can be misleading, as we saw in the case of Schwarzschild coordinates, that leads one to believe a particle would take infinite time to reach the event horizon. The calculation of the proper time of a radial infall showed us that not only does the particle reach the horizon in finite proper time, but that it also has no trouble crossing it. This fact was elucidated by switching to the Kruskal coordinates, where the $T-R$ diagram made clear that any particle or light can cross the event horizon. A crucial aspect of the coordinate transformations (\ref{eq:kruskal_coords_R}) that leads from spherical to Kruskal coordinates is the presence of a singularity at $r = 2M$, which is the reason why the new Kruskal line element presents only the physical singularity at $r = 0$. In the following chapters we will see how something similar can happen in the context of conformal transformations and singularities of the Ricci scalar. Notice that one is free to begin in Kruskal coordinates, where only the physical singularity is present, and then move to Schwarzschild's by means of a coordinate transformation that will introduce the coordinate singularity at $r = 2M$. However, it is more natural to begin with the Schwarzschild line element, since the spherical coordinates are familiar and more intuitive. A parallel to this will be drawn when we discuss the EF-JF physical equivalence in Chapter \ref{chapter5}.
\end{chapter}
\begin{chapter}{Conformal Transformations}
\label{chapter4}

\hspace{5mm}
Conformal transformations (CTs) are very useful in the context of gravitation, since they allow us to visualize global properties and the causal structure of spacetimes that can, in principle, be very complex. They also play an important role in $f(R)$ gravitation, since their use allows us to rewrite the Lagrangian in a familiar way that resembles the Einstein-Hilbert action \cite{sotiriou2010f}, as we will see.
\section{Definition and mathematical properties}
Consider a manifold with  metrics defined on it. These metrics are said to be conformal to each other if
\begin{equation}
    \Tilde{g}_{\mu\nu} = \Omega^2(\boldsymbol{x}) g_{\mu\nu},
    \label{eq:tc}
\end{equation}
where $\boldsymbol{x}$ represents the set of spacetime coordinates and $\Omega(\boldsymbol{x})$ is a smooth and non-vanishing function, called the conformal factor. CTs alter distances between two points in spacetime while preserving the angle between two vectors $\Tilde{X} = \Omega X$ and $\Tilde{Y} = \Omega Y$, which is given by
\begin{equation}
    \cos(\Tilde{X}, \Tilde{Y}) = \frac{\Tilde{g}_{\mu\nu}\tilde{X}^\mu \tilde{Y}^\nu}{\sqrt{|\Tilde{g}_{\alpha\beta}\tilde{X}^\alpha \tilde{X}^\beta|}\sqrt{|\Tilde{g}_{\gamma\delta}\tilde{Y}^\gamma \tilde{Y}^\delta|}}.
    \label{eq:angle1}
\end{equation}
Relation (\ref{eq:tc}) then gives us
\begin{align}
\cos(\Tilde{X}, \Tilde{Y}) &= \frac{\Tilde{g}_{\mu\nu}\tilde{X}^\mu \tilde{Y}^\nu}{\sqrt{|\Tilde{g}_{\rho\sigma}\tilde{X}^\rho \tilde{X}^\sigma|}\sqrt{|\Tilde{g}_{\lambda\tau}\tilde{Y}^\lambda \tilde{Y}^\tau|}} \notag \\
&= \frac{\Omega^4 g_{\mu\nu}X^\mu Y^\nu}{\sqrt{|\Omega^4 g_{\rho\sigma}X^\rho X^\sigma|}\sqrt{|\Omega^4 g_{\lambda\tau}Y^\lambda Y^\tau|}} \\
&= \frac{g_{\mu\nu}X^\mu Y^\nu}{\sqrt{|g_{\rho\sigma}X^\rho X^\sigma|}\sqrt{|g_{\lambda\tau}Y^\lambda Y^\tau|}} \notag \\
&= \cos(X,Y), \notag
\end{align}
which shows that the angle between vectors remains unaltered, since the conformal factor cancels out. This property is of great importance because since angles are left unaltered, the structure of light cones is left unchanged, preserving the causal structure of spacetime.

A crucial difference between conformal and coordinate transformations is that the line element is not invariant under the former, since the new metric is scaled by a function of the spacetime coordinates. This distinction can be exemplified if we consider the following static and spherically symmetric metrics, which are conformally related by $\phi = e^{-\sqrt{2\kappa/3}\tilde{\phi}(\tilde{r})}$ \cite{pretel2023compact}:
\begin{align}
    &d s^2 = -e^{\nu(r)} dt^2 + e^{\lambda(r)}dr^2 + r^2 d\Omega^2, \label{eq:ds_j} \\
    &d \tilde{s}^2 = -e^{\tilde{\nu}(\tilde{r})}dt^2 + e^{\tilde{\lambda}(\tilde{r})}d\tilde{r}^2 + \tilde{r}^2 d\Omega^{2}, \label{eq:ds_e}
\end{align}
the fact that $d\tilde{s}^2 = \phi\, ds^2$ leads us to
\begin{align}
    &\tilde{r}^2 = \phi \, r^2, \label{eq:r_phi_r} \\
    &e^{2\tilde{\nu}(\tilde{r})} = \phi \, e^{2\nu(r)}, \\
    &e^{2\tilde{\lambda}(\tilde{r})} =\phi \, e^{2\lambda(r)} \left(\frac{dr}{d\tilde{r}}\right)^2, \label{eq:lambda_phi_lambda}
\end{align}
and (\ref{eq:r_phi_r}) implies that
\begin{align}
    \frac{dr}{d\tilde{r}} &= \frac{d}{d\tilde{r}}(e^{\sqrt{\frac{k}{6}}\tilde{\phi}}\tilde{r}) \notag \\
    &= \phi^{-1/2}\left(\sqrt{\frac{k}{6}}\tilde{\phi}'(\tilde{r})\tilde{r} + 1\right).
    \label{eq:dr/dr}
\end{align}
The substitution of (\ref{eq:dr/dr}) into (\ref{eq:lambda_phi_lambda}) gives us
\begin{equation}
    e^{2\tilde{\lambda}(\tilde{r})} = e^{2\lambda(r)}\left(\sqrt{\frac{k}{6}}\tilde{\phi}'(\tilde{r})\tilde{r} + 1\right)^2,
\end{equation}
which shows how the radial contribution to (\ref{eq:ds_e}) changes compared to that of (\ref{eq:ds_j}).

A classic example of the use of CTs in physics consists on mapping points of a circle in the plane $xy$ into points of the complex plane $uv$ in order to find the potential inside a long hollow conducting cylinder \cite{brown2009complex}. This cylinder is split lengthwise, being separated in equal parts, as shown in the left side of figure \ref{fig:mapping} (from reference \cite{brown2009complex}), the upper and lower halves are then connected by using an insulating material. The upper half is grounded at zero potential, while the lower half is held at a fixed potential. We then consider a cross section of the cylinder, with the objective being to find the potential $V(x,y)$ between the halves: a harmonic function which satisfies Laplace's equation, $\nabla^2 V(x,y) = 0$, inside the circle $x^2 + y^2 = 1$. 
\begin{figure}[H]
    \centering
    \includegraphics{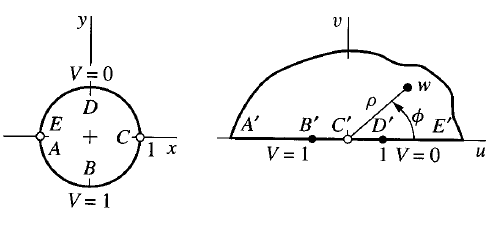}
    \caption{On the left panel we have a cross section of the conductor which has been split in half lengthwise, with the halves being connected by an insulating material. The upper half is grounded while the lower half is held at a fixed potential. On the right panel we have the result of the conformal mapping generated by (\ref{eq:mapping1}), which maps points inside the unitary circle in the $xy$ plane, associated to the number $z = x +iy$, into points of the upper half of the plane $uv$, associated to the number $w = u + iv$. This figure was obtained from \cite{brown2009complex}.}
    \label{fig:mapping}
\end{figure}
The conformal transformation which maps the interior of the circle in the upper half of the complex plane, the upper half of the circle in the positive real axis of that plane and the lower half in the negative real axis of the plane is given by \cite{brown2009complex}
\begin{equation}
    w = i \frac{1-z}{1+z},
    \label{eq:mapping1}
\end{equation}
in which $w = u + i v$ and $z = x + iy$, with $x$ and $y$ being the physical coordinates, see figure \ref{fig:mapping}. If we consider $w = \rho e^{i\phi}$, the polar form of $w$, and then consider the expression
\begin{equation}
   \frac{1}{\pi} \log{w} = \frac{1}{\pi}\log{\rho} +\frac{i}{\pi}\phi,
   \label{eq:mapping2}
\end{equation}
we notice that the imaginary part of (\ref{eq:mapping2}) is equal to 0 and 1 when $\phi = 0$ (positive real axis) and $\phi = \pi$ (negative real axis), respectively, satisfying the boundary conditions of the physical problem. This suggests that the harmonic function that satisfies the boundary conditions is
\begin{equation}
    V = \frac{1}{\pi}\arctan{\frac{u}{v}},
\end{equation}
which can be expressed in terms of the original coordinates $x$ and $y$, giving us
\begin{equation}
    V(x,y) = \frac{1}{\pi}\arctan\left({\frac{1-x^2-y^2}{2y}}\right).
\end{equation}
We have that $V(0,1) = 0$ and $V(0,-1) = 1$, satisfying the boundary condition. This and the fact that $V(x,y)$ is a harmonic function, implies that it is the unique solution to our problem, as stated by the uniqueness theorem of electrostatics \cite{griffiths2021introduction}.

\section{Conformal Transformations In $f(R)$ Theories}
\label{sec:conformal_in_f(R)}
We begin by applying a Legendre transformation to the $f(R)$ action, (\ref{eq:f(R)_action}), in order to replace $R$ by $\phi = df(R)/dR$ as the main variable. This leads us to
\begin{align}
    &S_{JF} = \frac{1}{2\kappa}\int{d^4x \sqrt{-g}\left[\phi R - U(\phi)\right]} + S_m,
     \label{eq:JFaction}\\
    &U(\phi) = R(\phi)\phi - f(R(\phi)), \label{eq:JF_potential}
\end{align}
where the Legendre transformation is given by equation (\ref{eq:JF_potential}), in which $U(\phi)$ acts as a potential associated to the field $\phi$ and (\ref{eq:JFaction}) is called the JF action. The associated field equations can be obtained, as before, by taking the variation with respect to $g^{\mu\nu}$ and $\phi$ or simply by rewriting equation (\ref{eq:f(R)_equation}) in terms of $\phi$ and $U(\phi)$, which gives us
\begin{align}
&G_{\mu\nu} = \frac{\kappa}{\phi} T_{\mu\nu}^{(m)} + \kappa T_{\mu\nu}^{(\phi)},
\label{eq:JFequation1dummy} \\
&R = U'(\phi), \\
&T_{\mu\nu}^{(\phi)} \equiv \frac{1}{2\kappa\phi} g_{\mu\nu} U(\phi) - \frac{1}{\phi} (\nabla_\mu \nabla_\nu \phi - g_{\mu\nu} \square \phi).
\label{eq:JFequation2dummy}
\end{align}
Notice, however, that in this form, equation (\ref{eq:JFequation1dummy}) presents an effective gravitational constant proportional to $\kappa/\phi$, and $T_{\mu\nu}^{(m)}$ becomes a source of the field $\phi$. This can be easily seen by taking the divergence of (\ref{eq:JFequation1dummy}), which leads to
\begin{equation}
\nabla^{\mu}T_{\mu\nu}^{(\phi)} = -\kappa T_{\mu\nu}^{(m)} \frac{\partial^{\mu}\phi}{\phi^2},
\end{equation}
where we used that $\nabla^{\mu}G_{\mu\nu} = 0$, because of the Bianchi identities and $\nabla^{\mu}T_{\mu\nu}^{(m)} = 0$, since energy-momentum conservation is assumed to be true in the JF. Here we see clearly that the right-hand side, proportional to $T_{\mu\nu}^{(m)}$, works as a source for $T_{\mu\nu}^{(\phi)}$. This is inconsistent with the fact that $T_{\mu\nu}^{(m)}$ is not affected by $T_{\mu\nu}^{(\phi)}$ since $T_{\mu\nu}^{(m)}$ is conserved by itself, as indicated by $\nabla^{\mu}T_{\mu\nu}^{(m)} = 0$.

This contradiction can be avoided if we multiply both sides of (\ref{eq:JFequation1dummy}) by $\phi$ and then add $G_{\mu\nu}$ to each side, leading us to the correct definition of $T_{\mu\nu}^{(\phi)}$:
\begin{align}
    &G_{\mu\nu} = \kappa\left(T_{\mu\nu}^{(m)} + T_{\mu\nu}^{(\phi)}\right) \label{eq:JFequation1}, \\
    &T_{\mu\nu}^{(\phi)} = \frac{1}{\kappa}\left[\nabla_{\mu}\nabla_{\nu}\phi - g_{\mu\nu}\left(\frac{1}{2}U(\phi) + \square \phi\right) + \left(1-\phi\right)G_{\mu\nu}\right] \label{eq:JF_Tmunu_phi}.
\end{align}
Notice that, in this case, the divergence of (\ref{eq:JFequation1}) leads to 
\begin{equation}
    \nabla^{\mu}T_{\mu\nu}^{(\phi)} = 0,
\end{equation}
and $T_{\mu\nu}^{(m)}$ does not work as a source of $\phi$, as expected in the JF. Furthermore, taking the trace of (\ref{eq:JFequation1}) and using (\ref{eq:JFequation2dummy}) to write $R$ in terms of $U'(\phi)$ leads us to the equation for the dynamics of the field $\phi$:
\begin{equation}
    \square \phi + \frac{2}{3}U(\phi) - \frac{\phi}{3}\frac{dU(\phi)}{d\phi} = \frac{1}{3}\kappa T^{(m)}.
    \label{eq:JF_phi_equation}
\end{equation}
Notice that this is still a higher-order differential equation for the metric, since $\phi = d f/ d R$. Therefore, if $\phi = 1$, we recover GR.

Now we will apply a conformal transformation to the metric in order to rewrite the action in what is called the Einstein frame representation. The CT is
\begin{align}
    &\Tilde{g}_{\mu\nu} = f'(R) g_{\mu\nu} = \phi \, g_{\mu\nu},
    \label{eq:CT} \\
    &\tilde{\phi} \equiv -\sqrt{\frac{3}{2\kappa}}\ln \phi,
    \label{eq:my_ct}
\end{align}
where $\tilde{\phi}$ is called the conformal field. Notice that the condition $f'>0$ mentioned in Chapter \ref{chapter2} is necessary, since the argument of $\ln$ needs to be positive if we want $\tilde{\phi}$ to not be complex. It is also convenient, since $\phi < 0$ would induce a change in the signature of $\tilde{g}_{\mu\nu}$. The Ricci scalar and the covariant derivative transform as shown in references \cite{carroll2019spacetime} and \cite{wald2010general} respectively, where the CT is written as $\Tilde{g}_{\mu\nu} = \omega^2(x)g_{\mu\nu}$. We then identify $\omega^2 = \phi = e^{-\sqrt{\frac{2k}{3}}\Tilde{\phi}}$, which leads us to
\begin{align}
    R &= \omega^2 \Tilde{R} + 2(n-1)\Tilde{g}^{\alpha \beta} \omega (\Tilde{\nabla}_{\alpha} \Tilde{\nabla}_{\beta}\omega) - n(n-1)\Tilde{g}^{\alpha \beta}(\Tilde{\nabla}_{\alpha}\omega)(\Tilde{\nabla}_{\beta}\omega) \notag \\ 
    &= \phi \Tilde{R} + 6\Tilde{g}^{\alpha \beta}\sqrt{\phi}\underbrace{(\Tilde{\nabla}_{\alpha} \Tilde{\nabla}_{\beta}\sqrt{\phi})}_{I} - 12\Tilde{g}^{\alpha \beta}\underbrace{(\Tilde{\nabla}_{\alpha}\sqrt{\phi})(\Tilde{\nabla}_{\beta}\sqrt{\phi})}_{II},
    \label{eq:ricci_CT} \\
    \tilde{\nabla}_{\alpha} V_{\beta} & = \nabla_{\alpha} V_{\beta} + \left(g_{\alpha\beta}g^{\mu\nu}\nabla_{\nu}\ln \sqrt{\phi} - 2 \delta_{(\alpha}^{\mu} \nabla_{\beta)} \ln{\sqrt{\phi}} \right) V_{\mu}\label{eq:CT_nabla},
\end{align}
where $\tilde{R}$ and $\tilde{\nabla}_{\mu}$ are the Ricci scalar and covariant derivative associated to $\tilde{g}_{\mu\nu}$, and $V_{\mu}$ is a dual vector. The brackets on the indices of equation (\ref{eq:CT_nabla}) denote the symmetrization operation. Notice that for a scalar quantity like $\phi$ we have $\tilde{\nabla}_{\alpha}\phi = \nabla_{\alpha} \phi = \partial_{\alpha} \phi$.

We then write $\phi$ in terms of $\Tilde{\phi}$ using equation (\ref{eq:my_ct}). The term $I$ becomes:
\begin{align}
    (\Tilde{\nabla}_{\alpha} \Tilde{\nabla}_{\beta}\sqrt{\phi}) &= \Tilde{\nabla}_{\alpha}(-\frac{1}{2}\sqrt{\frac{2k}{3}}e^{-\frac{1}{2}\sqrt{\frac{2k}{3}}\Tilde{\phi}}\Tilde{\nabla}_{\beta}\Tilde{\phi}) \notag \\ 
    &= -\frac{1}{2}\sqrt{\frac{2k}{3}}\left[-\frac{1}{2}\sqrt{\frac{2k}{3}}e^{-\frac{1}{2}\sqrt{\frac{2k}{3}}\Tilde{\phi}}(\Tilde{\nabla}_{\alpha}\Tilde{\phi})(\Tilde{\nabla}_{\beta}\Tilde{\phi}) + e^{-\frac{1}{2}\sqrt{\frac{2k}{3}}\Tilde{\phi}}\Tilde{\nabla}_{\alpha}\Tilde{\nabla}_{\beta}\Tilde{\phi} \right] \notag \\
    &= -\frac{1}{2}\sqrt{\frac{2k}{3}}e^{-\frac{1}{2}\sqrt{\frac{2k}{3}}\Tilde{\phi}}\left[-\frac{1}{2}\sqrt{\frac{2k}{3}}(\Tilde{\nabla}_{\alpha}\Tilde{\phi})(\Tilde{\nabla}_{\beta}\Tilde{\phi}) + \Tilde{\nabla}_{\alpha}\Tilde{\nabla}_{\beta}\Tilde{\phi} \right],
\end{align}
as for the term $II$:
\begin{align}
    (\Tilde{\nabla}_{\alpha}\sqrt{\phi})(\Tilde{\nabla}_{\beta}\sqrt{\phi}) &= (-\frac{1}{2}\sqrt{\frac{2k}{3}}e^{-\frac{1}{2}\sqrt{\frac{2k}{3}}\Tilde{\phi}}\Tilde{\nabla}_{\alpha}\Tilde{\phi})(-\frac{1}{2}\sqrt{\frac{2k}{3}}e^{-\frac{1}{2}\sqrt{\frac{2k}{3}}\Tilde{\phi}}\Tilde{\nabla}_{\beta}\Tilde{\phi}) \notag \\
    &= \frac{k}{6}e^{-\sqrt{\frac{2k}{3}}\Tilde{\phi}}(\Tilde{\nabla}_{\alpha}\Tilde{\phi})(\Tilde{\nabla}_{\beta}\Tilde{\phi}).
\end{align}
Substituting terms $I$ and $II$ in equation (\ref{eq:ricci_CT}) yields
\begin{align}
    R &= \phi \Tilde{R} + 6\Tilde{g}^{\alpha\beta}e^{-\frac{1}{2}\sqrt{\frac{2k}{3}}\Tilde{\phi}}-\frac{1}{2}\sqrt{\frac{2k}{3}}e^{-\frac{1}{2}\sqrt{\frac{2k}{3}}\Tilde{\phi}}\left[-\frac{1}{2}\sqrt{\frac{2k}{3}}(\Tilde{\nabla}_{\alpha}\Tilde{\phi})(\Tilde{\nabla}_{\beta}\Tilde{\phi}) + \Tilde{\nabla}_{\alpha}\Tilde{\nabla}_{\beta}\Tilde{\phi} \right] \notag \\
    &- 12\Tilde{g}^{\alpha\beta}\frac{k}{6}e^{-\sqrt{\frac{2k}{3}}\Tilde{\phi}}(\Tilde{\nabla}_{\alpha}\Tilde{\phi})(\Tilde{\nabla}_{\beta}\Tilde{\phi}) \notag \\
    &= \phi \Tilde{R} - k\phi(\Tilde{\nabla}^{\alpha}\Tilde{\phi})(\Tilde{\nabla}_{\beta}\Tilde{\phi}) - 3\sqrt{\frac{2k}{3}}\phi\Tilde{\nabla}^{\beta}\Tilde{\nabla}_{\beta}\Tilde{\phi}
    \label{eq:R(tilde_R)},
\end{align}
to rewrite $\sqrt{-g}$ we recall that $\Tilde{g}_{\alpha\beta} = \phi \, g_{\alpha\beta} \implies \Tilde{g} = \phi^4 g \implies \sqrt{-\Tilde{g}} = \phi^2\sqrt{-g}$. Plugging these results in the action (\ref{eq:JFaction}) leads to
\begin{align}
    S_{EF} &= \frac{1}{2k}\int{d^4x \sqrt{-g}\left[\phi\left(\phi \Tilde{R} - k\phi(\Tilde{\nabla}^{\alpha}\Tilde{\phi})(\Tilde{\nabla}_{\beta}\Tilde{\phi}) - 3\sqrt{\frac{2k}{3}}\phi\underbrace{\Tilde{\nabla}^{\beta}\Tilde{\nabla}_{\beta}\Tilde{\phi}}_{S.T}\right) - U\right]} + S_m \notag \\
    &= \int{d^4x \sqrt{-\Tilde{g}}\left[\frac{\Tilde{R}}{2k} - \frac{1}{2}\Tilde{g}^{\mu\nu}(\Tilde{\nabla}_{\mu} \Tilde{\phi})(\Tilde{\nabla}_{\nu}\Tilde{\phi}) - V(\Tilde{\phi})\right]} + S_m\left(e^{\sqrt{\frac{2\kappa}{3}}\tilde{\phi}} \Tilde{g}_{\mu\nu}, \psi\right),
    \label{eq:EFaction}
\end{align}
where the surface term $S.T$ term is equivalent to a surface integral by Stokes' theorem, which, as done before, we set to zero at infinity and the new potential is $V(\Tilde{\phi}) = U/2k\phi^2$. In this form the action is said to be written in the Einstein Frame (EF) since it has the form of the standard GR action with a propagating scalar field and a potential associated to it. Notice that in the EF there is a coupling between matter and the conformal field $\Tilde{\phi}$ via the factor $e^{\sqrt{\frac{2\kappa}{3}}\tilde{\phi}}$ in the matter action $S_m$, so one must not presume that in this frame everything will be the same as in GR, which really is not true as will be shown in the next section.

The field equations are obtained, as usual, by taking the variation of (\ref{eq:EFaction}) with respect to $\Tilde{g}^{\mu\nu}$ and $\Tilde{\phi}$. Beginning with the former, gives us
\begin{equation}
    \delta S_{EF} = \int{d^4x \left[\underbrace{\delta(\sqrt{-\Tilde{g}}\tilde{g}^{\mu\nu}\Tilde{R}_{\mu\nu})}_I -\underbrace{\delta(\frac{1}{2}\sqrt{-\Tilde{g}}\Tilde{g}^{\mu\nu}(\Tilde{\nabla}_{\mu} \Tilde{\phi})(\Tilde{\nabla}_{\nu}\Tilde{\phi}))}_{II} - \underbrace{\delta(\sqrt{-\Tilde{g}}V(\Tilde{\phi}))}_{III} \right]}.
    \label{eq:EF_variation}
\end{equation}
The term $I$ is the standard Einstein-Hilbert action and has already been computed in Chapter \ref{chapter2}, in the process of obtaining the field equations of GR:
\begin{equation}
    I: \delta(\sqrt{-\Tilde{g}}\tilde{g}^{\mu\nu}\Tilde{R}_{\mu\nu}) = \sqrt{-\Tilde{g}}(\Tilde{R}_{\mu\nu} - \frac{1}{2}\Tilde{R}\Tilde{g}_{\mu\nu})\delta \Tilde{g}^{\mu\nu},
    \label{eq:EF_variation_I}
\end{equation}
the term $II$ becomes
\begin{align}
 II:  \delta(\frac{1}{2}\sqrt{-\Tilde{g}}\Tilde{g}^{\mu\nu}(\Tilde{\nabla}_{\mu} \Tilde{\phi})(\Tilde{\nabla}_{\nu}\Tilde{\phi})) &= \tilde{\nabla}_{\mu}\tilde{\phi}\tilde{\nabla}_{\nu}\tilde{\phi}(\tilde{g}^{\mu\nu}\delta \sqrt{-\tilde{g}} + \sqrt{-\tilde{g}}\delta \tilde{g}^{\mu\nu}) \notag \\
 &= (1 - \frac{1}{2}\sqrt{-\tilde{g}}\Tilde{g}_{\alpha\beta}\tilde{g}^{\mu\nu})\Tilde{\nabla}_{\mu}\tilde{\phi}\tilde{\nabla}_{\nu}\tilde{\phi}\delta \tilde{g}^{\alpha\beta},
 \label{eq:EF_variation_II}
\end{align}
where we used equation (\ref{eq:sqrt_variation}) to express $\delta\sqrt{-\tilde{g}}$ in terms of $\delta \tilde{g}^{\mu\nu}$. Term $III$ is just
\begin{equation}
    III: \delta(\sqrt{-\tilde{g}}V(\tilde{\phi})) = -\frac{1}{2}\sqrt{\Tilde{g}}\,\Tilde{g}_{\mu\nu}V(\tilde{\phi})\delta \tilde{g}^{\mu\nu}
    \label{eq:EF_variation_III}.
\end{equation}
Substituting equations (\ref{eq:EF_variation_I}), (\ref{eq:EF_variation_II}) and (\ref{eq:EF_variation_III}) into (\ref{eq:EF_variation}) and using the least action principle leads us to
\begin{align}
    \delta S_{EF} = \int d^4x\,\sqrt{-\tilde{g}}\Big[\frac{1}{2\kappa}(\Tilde{R}_{\mu\nu} - \frac{1}{2}\tilde{R}\tilde{g}_{\mu\nu}) &+ \frac{1}{4}\tilde{g}_{\mu\nu}\tilde{g}^{\alpha\beta}\tilde{\nabla}_{\alpha}\tilde{\phi}\tilde{\nabla}_{\beta}\tilde{\phi} \notag \\ &+\tilde{\nabla}_{\mu}\tilde{\phi}\tilde{\nabla}_{\nu}\tilde{\phi} + \frac{1}{2}\tilde{g}_{\mu\nu}V(\tilde{\phi})\Big]\delta \tilde{g}^{\mu\nu}, 
\end{align}
and the least action principle leads us to
\begin{align}
    &\tilde{G}_{\mu\nu} = \kappa(\tilde{T}_{\mu\nu}^{(m)} + \tilde{T}_{\mu\nu}^{(\tilde{\phi})}), \label{eq:EF_field_eq_metric} \\
    &\tilde{T}_{\mu\nu}^{\tilde{(\phi)}} = \tilde{\nabla}_{\mu}\tilde{\phi}\tilde{\nabla}_{\nu}\tilde{\phi} - \frac{1}{2}\tilde{g}_{\mu\nu}\tilde{g}^{\alpha\beta}\tilde{\nabla}_{\alpha}\tilde{\phi}\tilde{\nabla}_{\beta}\tilde{\phi} - \tilde{g}_{\mu\nu}V(\tilde{\phi}), \label{eq:EF_phi_momentum_tensor} \\
    &\tilde{T}_{\mu\nu}^{(m)} = \frac{-2}{\sqrt{-\tilde{g}}}\frac{\delta S_m}{\delta\tilde{g}^{\mu\nu}}. \label{eq:EF_Tm_tensor}
\end{align}
This is simply the standard Einstein equation with the presence of a scalar field $\tilde{\phi}$ and an associated potential $V(\tilde{\phi})$, as is expected. Now we take the variation of the EF action with respect to $\tilde{\phi}$:
\begin{equation}
    \delta S_{EF} = \int{d^4x \, \sqrt{-\tilde{g}}\left[-\frac{1}{2}\delta(\tilde{g}^{\alpha\beta}\tilde{\nabla}_{\alpha}\tilde{\phi}\tilde{\nabla}_{\beta}\tilde{\phi}) - \frac{dV}{d\tilde{\phi}}\delta \tilde{\phi}\right] } + \delta S_m.
    \label{eq:EF_phi_variation}
\end{equation}
Notice that the term of (\ref{eq:EFaction}) proportional to $\tilde{R}$ vanishes and $\delta$ does not act on $\tilde{g}_{\alpha\beta}$ for the same reason: the metric and the field are independent entities. In this frame this can be a source of confusion because, when switching frames, we write said quantities in terms of their counterparts from the other frame, that is, we write $R$ in terms of $\tilde{R}$ and $\tilde{\phi}$, and vice-versa, like in equation (\ref{eq:R(tilde_R)}), and the same goes for the metric. Although these relations do exist, they should only be used to switch frames, but once this is done, the metric and the field are treated as separate entities, as opposed to what happens in the JF, where the scalar field absorbs the extra degrees of freedom from the metric. Getting back to the calculation, the first term inside the square brackets of (\ref{eq:EF_phi_variation}) becomes
\begin{align}
\delta(\tilde{g}^{\alpha\beta}\tilde{\nabla}_{\alpha}\tilde{\phi}\tilde{\nabla}_{\beta}\tilde{\phi}) &= \tilde{g}^{\alpha\beta}\left[\delta(\tilde{\nabla}_{\alpha}\tilde{\phi})\tilde{\nabla}_{\beta}\tilde{\phi} + \delta(\tilde{\nabla}_{\beta}\tilde{\phi})\tilde{\nabla}_{\alpha}\tilde{\phi}\right] \notag \\
&= 2\tilde{g}^{\alpha\beta}\tilde{\nabla}_{\alpha}(\delta \tilde{\phi})\tilde{\nabla}_{\beta}\tilde{\phi},
\label{eq:delta_g_nabla_nabla_phi1}
\end{align}
where we used the fact that $\delta$ commutes with $\nabla$ because the variation is being taken with respect to $\tilde{\phi}$ and that the indices $\alpha$ and $\beta$ can switch places in the second term inside the square brackets. This can be integrated by parts since
\begin{equation}
    \tilde{\nabla}_{\alpha}(\tilde{g}^{\alpha\beta}\delta \tilde{\phi}\tilde{\nabla}_{\beta}\tilde{\phi}) = \tilde{g}^{\alpha\beta}\tilde{\nabla}_{\alpha}(\delta \tilde{\phi})\tilde{\nabla}_{\beta}\tilde{\phi} + \delta\tilde{\phi}\tilde{g}^{\alpha\beta}\tilde{\nabla}_{\alpha}\tilde{\nabla}_{\beta}\tilde{\phi},
    \label{eq:delta_g_nabla_nabla_phi2}
\end{equation}
where the integral of the left-hand side of (\ref{eq:delta_g_nabla_nabla_phi2}) is equivalent to a surface integral which, as usual, we set to zero at infinity. Using (\ref{eq:delta_g_nabla_nabla_phi1}) and (\ref{eq:delta_g_nabla_nabla_phi2}) in (\ref{eq:EF_phi_variation}) and using, once again, the least action principle gives us
\begin{equation}
   \Tilde{g}^{\alpha\beta}\tilde{\nabla}_{\alpha}\tilde{\nabla}_{\beta}\tilde{\phi} - \frac{dV}{d\tilde{\phi}} + \frac{1}{\sqrt{-\tilde{g}}}\frac{\delta S_m}{\delta\tilde{\phi}} = 0,
   \label{eq:EF_field_eq_dummy}
\end{equation}
and the $\delta S_m/\delta\tilde{\phi}$ term becomes
\begin{align}
    \frac{1}{\sqrt{-\tilde{g}}}\frac{\delta S_m}{\delta\tilde{\phi}} &= \frac{\delta S_m}{\delta\tilde{g}_{\mu\nu}}\frac{d\tilde{g}_{\mu\nu}}{d\tilde{\phi}} \notag \\
    & = \sqrt{\frac{\kappa}{6}}\frac{-2}{\sqrt{-\tilde{g}}}\frac{\delta S_m}{\delta\tilde{g}_{\mu\nu}}\tilde{g}_{\mu\nu} \notag \\
    &= \sqrt{\frac{\kappa}{6}}\tilde{T}_{(m)}^{\mu\nu}\tilde{g}_{\mu\nu}.
    \label{eq:EF_Tm_tensor_dummy},
\end{align}
where we used relation (\ref{eq:CT}) to write $\tilde{g}_{\mu\nu}$ in terms of $g_{\mu\nu}$. Finally, plugging (\ref{eq:EF_Tm_tensor_dummy}) into (\ref{eq:EF_field_eq_dummy}) gives us 
\begin{equation}
    \square\tilde{\phi} - \frac{d V(\tilde{\phi})}{d \tilde{\phi}} + \sqrt{\frac{\kappa}{6}}\tilde{T}_{(m)}^{\mu\nu}\tilde{g}_{\mu\nu} = 0,
    \label{eq:EF_phi_equation}
\end{equation}
which describes the dynamics of the field $\tilde{\phi}$.

\section{$f(R)$ and Brans-Dicke Theory}
The Brans-Dicke (BD) action in the JF for a scalar field $\Phi$ with a potential $U(\Phi)$ is \cite{faraoni2011beyond}
\begin{equation}
    S_{BD} = \frac{1}{2\kappa}\int{d^4x \, \sqrt{-g}\left[\Phi R - \frac{\omega}{\Phi} g^{\mu\nu} \nabla_{\mu} \Phi \nabla_{\nu} \Phi - U(\Phi) \right]} + S_m(g_{\mu\nu},\psi),
    \label{eq:BD_action}
\end{equation}
in which $\omega$ is a dimensionless Brans-Dicke parameter. The field equations of the theory are obtained by the usual method of taking the variation of the action with respect to $g^{\mu\nu}$ and $\Phi$, which yields
\begin{align}
   &G_{\mu\nu} = \frac{\kappa}{\Phi}T_{\mu\nu}^{(m)} + \frac{\omega}{\Phi^2}\left(\nabla_{\mu} \Phi \nabla_{\nu} \Phi - \frac{1}{2}g_{\mu\nu} \nabla^{\alpha} \Phi \nabla_{\alpha} \Phi\right) + \frac{1}{\Phi}\left( \nabla_{\mu} \nabla_{\nu} \Phi - g_{\mu\nu} \square \Phi\right) - \frac{U(\Phi)}{2\Phi} g_{\mu\nu}, \\
   &R - \frac{\omega}{\Phi^2} \nabla^{\alpha}\Phi \nabla_{\alpha}\Phi - \frac{dU(\Phi)}{d\Phi} + \frac{2\omega}{\Phi} \square\Phi = 0,
\end{align}
and we see that, when $\omega = 0$, these equations reduce to (\ref{eq:JFequation1dummy}) and (\ref{eq:JFequation2dummy}) with $\Phi = \phi$, which might suggest $f(R)$ theories are a particular case of the BD theory. Notice, however, that in BD the field $\Phi$ is a fundamental entity, while in $f(R)$ it is an auxiliary field defined as $f'(R)$. Also, in its original formulation \cite{brans1961mach}, the BD action did not include a potential, which arises naturally in the JF representation of the $f(R)$ theory, which in turn has no kinetic term analogous to the $\omega$ contribution in (\ref{eq:BD_action}). For these reasons, the experimental bound $\omega > 40000$ \cite{perivolaropoulos2010ppn} does not imply that $f(R)$ theories should be discarded, since they are Brans-Dicke-like, but not actually a particular case of the BD theory.

\section{The Einstein And Jordan Frames: Which One Is Physical? }
\label{physicalframes}
We have seen how the JF and EF are equivalent mathematical representations of the same theory, but are they also physically equivalent? First we need to understand what it means to be physically equivalent. If this is true, then predictions made in both frames must comply, in the sense that they should describe the same physical reality and that physical experiments would yield the same qualitative results in either frame.

In reference \cite{faraoni2007pseudo} the authors address the matter of the physical equivalence of the JF and EF, focusing on the latter's adoption of "running units" (see subsection \ref{subsec:running_units}), a concept in which the units of measurement change with the scalar field through the conformal factor $\Omega$. This approach maintains conformal invariance in the EF, meaning that the form of physical laws remains unchanged when units are rescaled with the appropriate power of $\Omega$.

Although the conservation equation for the energy-momentum tensor and the geodesic equation for a free-falling particle in the EF acquire additional terms due to the coupling with the conformal field, this modification does not compromise the measurement of physical observables. The authors argue that in experimental practice, it is the ratio of some physical quantity $A$ to a chosen unit of measurement that is determined, rather than an absolute value. This ensures that the observed physics remains consistent, regardless of the apparent non-conservation introduced by the conformal field in the Einstein frame.
\subsection{Jordan Frame}
If we are to say that a frame is physical or not, it is surely important to analyze the energy-momentum conservation in it. In section \ref{sec:conformal_in_f(R)}, we have seen that the field $\phi$ couples minimally with matter, via the metric, which is necessary since the conservation of energy-momentum is taken as a fact. A direct consequence of this is that free-falling test particles in the JF will follow the geodesics of $g_{\mu\nu}$. In this frame, the following equations hold:
\begin{align}
    &\nabla^{\mu} T_{\mu\nu}^{(m)} = 0, \label{eq:JF_Tmunu_conservation}\\
    &u^{\mu}\nabla_{\mu}u^{\nu} = 0, \label{eq:JF_geodesic_equation}
\end{align}
where (\ref{eq:JF_geodesic_equation}) is the JF geodesic equation and $u^{\mu}$ is a four-velocity.
We have also seen that the field $\phi$ obeys the field equation (\ref{eq:JF_phi_equation}), which we can rewrite in terms of an effective potential:
\begin{align}
  &\square \phi - \frac{dU_{eff}(\phi)}{d\phi} = 0, \\
  &\frac{dU_{eff}(\phi)}{d\phi} \equiv \frac{1}{3}\left(\phi\frac{dU(\phi)}{d\phi} - 2U(\phi) + \kappa T^{(m)}\right).
\end{align}
This has the form of a Klein-Gordon equation, and we can define the mass $m(\phi)$ of the field as \cite{Faraoni_2009_scalar}
\begin{equation}
   m^2(\phi) \equiv \frac{d^2U_{eff}(\phi)}{d\phi^2} \Bigg|_{\phi^*} = \frac{1}{3}\left(\phi\frac{d^2U(\phi)}{d\phi^2} - \frac{dU(\phi)}{d\phi}\right) \Bigg|_{\phi^*},
   \label{eq:JF_phi_mass}
\end{equation}
where $\phi^*$ is the value that minimizes $U_{eff}$. Using the definition (\ref{eq:JF_potential}) of $U(\phi)$, this can be written in terms of $f(R)$ as
\begin{equation}
    m^2(R) = \frac{1}{3}\left(\frac{f'(R)}{f''(R)} - R\right) \Bigg|_{R^*}
    \label{eq:JF_phi_mass_f'},
\end{equation}
where $R^*$ is the value of $R$ corresponding to $\phi^*$.

\subsection{Einstein Frame}
In the EF, we have seen that the conformal field $\tilde{\phi}$ couples non-minimally to matter, and a direct consequence of this can be seen if we consider the conservation of $\tilde{T}_{\mu\nu}^{(m)}$. Taking the four-divergence of (\ref{eq:EF_field_eq_metric}) yields
\begin{equation}
    \tilde{\nabla}^{\mu}\tilde{T}_{\mu\nu}^{(m)} + \tilde{\square}\tilde{\phi}\tilde{\nabla}_{\nu}\tilde{\phi} - \tilde{\nabla}_{\nu}V(\tilde{\phi}) = 0,
\end{equation}
and using (\ref{eq:EF_phi_equation}) to substitute $\tilde{\square}$ gives
\begin{equation}
    \tilde{\nabla}^{\mu}\tilde{T}_{\mu\nu}^{(m)} = \sqrt{\frac{\kappa}{6}}\tilde{T}^{(m)}\tilde{\nabla}_{\nu}\tilde{\phi},
    \label{eq:EF_Tmunu_conservation}
\end{equation}
and we see that only massless fields, for which the trace $\tilde{T}^{(m)}$ of the energy-momentum tensor vanishes, are conserved. This equation explicitly shows how the conformal field $\tilde{\phi}$ acts as a source term for $\tilde{T}_{\mu\nu}^{(m)}$. A consequence of this is the fact that test particles will not follow geodesics of $\tilde{g}_{\mu\nu}$, as can be seen by considering the conformal transformation of (\ref{eq:JF_geodesic_equation}), which yields \cite{wald2010general}
\begin{align}
    \tilde{u}^{\mu}\tilde{\nabla}_{\mu}\tilde{u}^{\nu} &= \tilde{u}^{\nu}\tilde{u}^{\alpha}\tilde{\nabla}_{\alpha}\ln{\phi^{1/2}} + g^{\nu\beta}\tilde{\nabla}_{\beta}\ln{\phi^{1/2}} \notag \\
    &= -\sqrt{\frac{\kappa}{6}}\left[\tilde{u}^{\nu}\tilde{u}^{\alpha}\tilde{\nabla}_{\alpha}\tilde{\phi} + \tilde{g}^{\nu\beta}\tilde{\nabla}_{\beta}\tilde{\phi}\right]
    \label{eq:EF_geodesic_equation},
\end{align}
in which we see clearly how the coupling of the field $\tilde{\phi}$ to matter changes the geodesic equation.
As was done in the JF, the field equation (\ref{eq:EF_phi_equation}) for $\tilde{\phi}$ can be written in terms of an effective potential $V_{eff}(\tilde{\phi})$, but first we recall that this equation is written in terms of $\tilde{T}^{(m)} = \tilde{T}^{\mu\nu}_{(m)}\tilde{g}_{\mu\nu}$, and we have seen that it is not a conserved quantity in this frame, which is why it is convenient to express it in terms the JF energy-momentum tensor:
\begin{align}
   \tilde{T}_{(m)}^{\mu\nu} \tilde{g}_{\mu\nu} &= \frac{-2}{\sqrt{-\tilde{g}}} \frac{\delta S_m}{\delta \tilde{g}_{\mu\nu}} \phi g_{\mu\nu} \notag \\
   &=\frac{-2}{\phi^2 \sqrt{-g}} \frac{\delta S_m}{\delta g_{\alpha\beta}} \frac{\delta g_{\alpha\beta}}{\delta \tilde{g}_{\mu\nu}} \phi g_{\mu\nu} \notag \\
   &= \frac{-2}{\phi^2 \sqrt{-g}} \frac{\delta S_m}{\delta g_{\alpha\beta}} \frac{1}{\phi} \delta_{\mu}^{\alpha} \delta_{\nu}^{\beta} \phi g_{\mu\nu} \notag \\
   &= e^{\sqrt{\frac{8\kappa}{3}}\tilde{\phi}} T_{(m)}^{\mu\nu} g_{\mu\nu}.
   \label{eq:Te_in_terms_of_Tj}
\end{align}
Now we plug (\ref{eq:Te_in_terms_of_Tj}) into \ref{eq:EF_phi_equation}, leading us to
\begin{align}
    &\tilde{\square} \tilde{\phi} - \frac{d V_{eff}(\tilde{\phi})}{d \tilde{\phi}} = 0,
    \label{eq:EF_phi_equation_effective}\\
    &\frac{d V_{eff}(\tilde{\phi})}{d \tilde{\phi}} = \frac{d V(\tilde{\phi})}{d \tilde{\phi}} - \sqrt{\frac{\kappa}{6}} e^{\sqrt{\frac{8\kappa}{3}}\tilde{\phi}} T^{(m)},
    \label{eq:EF_potential_effective}
\end{align}
and, as before, the mass of the field, $\tilde{m}(\tilde{\phi})$, is defined as
\begin{align}
    \tilde{m}^2(\tilde{\phi}) &\equiv \frac{d^2 V_{eff}(\tilde{\phi})}{d \tilde{\phi}^2}\Bigg|_{\tilde{\phi}^*} = \left(\frac{d^2 V(\tilde{\phi})}{d \tilde{\phi}^2} - \frac{2\kappa}{3}e^{\sqrt{\frac{8\kappa}{3}}\tilde{\phi}}T^{(m)}\right)\Bigg|_{\tilde{\phi}^*},
    \label{eq:EF_phi_mass}
\end{align}
where $\tilde{\phi}^*$ is the value that minimizes $V_{eff}$. Remembering that $V(\Tilde{\phi}) = U/2k\phi^2$ and, again, using (\ref{eq:JF_potential}) yields
\begin{equation}
    \tilde{m}^2(R) = \frac{\kappa}{3 f'(R)}\left[\frac{1}{\kappa}\left(R - \frac{4 f(R)}{f'(R)} + \frac{f'(R)}{f''(R)}\right) - \frac{2 T^{(m)}}{f'(R)}\right]\Bigg|_{R^*},
    \label{eq:EF_phi_mass_f'}
\end{equation}
where $R^*$ is the value of $R$ corresponding to $\tilde{\phi}^*$.
Equations (\ref{eq:EF_phi_mass}) and (\ref{eq:EF_phi_mass_f'}) show how the coupling of the conformal field, $\tilde{\phi}$, causes its mass to depend on $T^{(m)}$, that is, the mass changes based on the energy-matter distribution in a certain region, affecting the range of the field. In the solar system scale, where this density is large, the field has increased mass, and thus smaller range, masking its existence and associated fifth-force effects that would violate the equivalence principle, which would lead to deviations from GR. In larger cosmological scales, where the energy-matter density is small, the field would be lighter, and thus it would be possible to identify deviations from GR. This is known as the chameleon effect \cite{khoury2004chameleon}.

We have seen how the coupling of $\tilde{\phi}$ to matter has an impact on the fundamental equation of energy conservation, giving rise to exotic phenomena, like the chameleon effect. Some would say that this is enough to label the JF as the physical one, even if both frames are mathematically equivalent representations of the theory. To better understand the physicality of the frames, we will now discuss the running of units in the EF.

\subsection{Running Units}
\label{subsec:running_units}
Until this point, when transforming from the JF to the EF, we have seen that quantities like the EF metric components (\ref{eq:r_phi_r}-\ref{eq:lambda_phi_lambda}) and field mass (\ref{eq:EF_phi_mass}) will now scale with the conformal field $\tilde{\phi}(\tilde{r})$, and the units of time, length and mass become a function of the spacetime point. This is called the running of units and is discussed by Dicke in \cite{dicke1962mach}, where the conformal transformation technique was introduced for the Brans-Dicke theory \cite{brans1961mach}. There, it is shown that since $\tilde{g}_{\mu\nu} = \phi g_{\mu\nu}$, we have that the time ($t$), length ($l$) and masses ($m$) of all forms of matter scale as 
\begin{align}
    &d\tilde{t} = \phi^{1/2} d t, \\
    &d \tilde{l} = \phi^{1/2} d l, \label{eq:dicke_dx_phi_dx} \\
    &\tilde{m} = \phi^{-1/2} m. \label{eq:dicke_m_phi_m}
\end{align}
Notice that (\ref{eq:dicke_m_phi_m}) applies only to masses that come from the matter part, $S_m$, of the EF action (\ref{eq:EFaction}), which is not the case for the masses $m(\phi)$ and $\tilde{m}(\tilde{\phi})$ of the fields $\phi$ and $\tilde{\phi}$, defined in (\ref{eq:JF_phi_mass}) and (\ref{eq:EF_phi_mass}). The mass $m(\tilde{\phi})$ does scale with $\tilde{\phi}$, but not as shown in (\ref{eq:dicke_m_phi_m}).

Consider now an experiment that measures the mass $m_p$ of a proton, which is measured with respect to some arbitrary mass unit $m_u$, that is, the result of the experiment will be $m_p/m_u$ and, in the EF, we would have
\begin{equation}
    \frac{\tilde{m}_p}{\tilde{m}_u} = \frac{\phi^{-1/2} m_p}{\phi^{-1/2} m_u} = \frac{m_p}{m_u},
\end{equation}
and we see that in this kind of experiment, in which ratios are measured, the result does not depend of the frame used. In this sense, in Dicke's viewpoint, the JF and the EF are physically equivalent. We have seen how the running of units is a natural consequence of the EF and JF being conformally related that happens due to the anomalous coupling of the conformal field to matter. In fact, we can rewrite equation (\ref{eq:EF_geodesic_equation}) in terms of mass, by using relation (\ref{eq:dicke_m_phi_m}), and, for convenience, write it in a more compact form in which the correction to the geodesic equation can be see as consequence purely of the variation of the mass $\tilde{m}$ in the three-space of a co-moving observer. This can be done by introducing the three-dimensional metric on the rest space that is orthogonal to the four velocity $\tilde{u}^{a}$ of the particle:
\begin{equation}
    \tilde{h}_{\mu\nu} \equiv \tilde{g}_{\mu\nu} + \tilde{u}_{\mu}\tilde{u}_{\nu}.
    \label{eq:EF_3D_metric}
\end{equation}
Notice that $\tilde{h}^{\mu}_{\nu}$ acts as a projector operator on the rest space of an observer. Finally, plugging (\ref{eq:EF_3D_metric}) and (\ref{eq:dicke_m_phi_m}) into (\ref{eq:EF_geodesic_equation}) leaves with
\begin{equation}
    \tilde{u}^{\mu} \tilde{\nabla}_{\mu} \tilde{u}^{\nu} = - \frac{\tilde{h}^{\nu\alpha} \tilde{\nabla}_{\alpha} \tilde{m}}{\tilde{m}},
    \label{eq:EF_geodesic_mass}
\end{equation}
in which we see explicitly how the change in the geodesic equation is caused by the gradient of the mass, which has a dependence on the spacetime point. We can also express this in terms of the unit mass, $\tilde{m}_{u}$, of the frame by noticing that the constancy of the ratio $\tilde{m}/\tilde{m}_{u}$ implies that
\begin{equation}
   \frac{\tilde{\nabla}_{\alpha} \tilde{m}}{\tilde{m}} = \frac{\tilde{\nabla}_{\alpha} \tilde{m}_{u}}{\tilde{m}_{u}},
\end{equation}
and so, equation (\ref{eq:EF_geodesic_mass}) becomes
\begin{equation}
    \tilde{u}^{\mu} \tilde{\nabla}_{\mu} \tilde{u}^{\nu} = - \frac{\tilde{h}^{\nu\alpha} \tilde{\nabla}_{\alpha} \tilde{m}_{u}}{\tilde{m}_{u}},
    \label{eq:EF_geodesic_mass_unit}
\end{equation}
where we see that the correction for the geodesic equation can be seen as being caused by the variation of the mass unit itself. 

Another instance of the importance to have running units in the EF emerges when we consider the requirement that physics be conformally invariant, in the sense that equations should remain in the same form after a conformal transformation. Consider, as an example, the equation that describes a conformally-coupled Klein-Gordon field in the JF, with non-zero mass \cite{grib1995problem}
\begin{equation}
  \square \psi - \frac{R}{6} \psi - m^2 \psi = 0.
  \label{eq:JF_klein_eq}
\end{equation}
If we rewrite this in terms of tilded quantities, it can be shown that \cite{faraoni2007pseudo}
\begin{equation}
    g^{\mu\nu}\nabla_{\mu}\nabla_{\nu}\psi - \frac{R}{6}\psi = \phi^{3/2}\left[\tilde{g}^{\mu\nu} \tilde{\nabla}_{\mu} \tilde{\nabla}_{\nu} \tilde{\psi} - \frac{\tilde{R}}{6}\tilde{\psi}\right],
\end{equation}
and the fact that $m$ scales as shown in (\ref{eq:dicke_m_phi_m}) and  $\tilde{\psi} = \phi^{-1/2} \psi$ \cite{grib1995problem} means that (\ref{eq:JF_klein_eq}) transforms to
\begin{equation}
  \tilde{\square} \tilde{\psi} - \frac{\tilde{R}}{6} \tilde{\psi} - \tilde{m}^2 \tilde{\psi} = 0,
  \label{eq:EF_klein_eq}
\end{equation}
and we see that the equations have the same form in both frames, which is why, in this sense, they are said to be conformally invariant, even though the individual components of the equation are not. Note that such invariance would not be possible if the mass unit did not scale as it does, showing, again, that the running of units is fundamental if we are to say that the EF is physical, as argued in \cite{faraoni2007pseudo}.

We have seen how CTs play an important role in the context of $f(R)$ gravitation, allowing us to represent the theory in two mathematically equivalent frames: Jordan's and Einstein's. The physicality of the frames was discussed and while not being obvious at first, we have seen that the running units of the EF allow physical laws to be conformally invariant, as well as experiments which measure ratios between some observable and its unit measure to yield results that agree in both frames.
\end{chapter}
\begin{chapter}{Conformal Transformations And Singularities Of The Ricci Scalar}
\label{chapter5}
In Chapter \ref{chapter4} we discussed the differences between the Einstein and Jordan frames. Here we will address the main topic of this dissertation: the behavior of the Ricci scalar under conformal transformations and how physical singularities can be affected by it.
\section{The Ricci Scalar} 
We begin with the expression for the conformal transformation of the Ricci scalar from the JF to the EF \cite{carroll2019spacetime}:
\begin{equation}
    \Tilde{R} = \frac{R}{f'} - \frac{6 g^{\alpha\beta}\nabla_{\alpha}\nabla_{\beta}f'^{1/2}}{f'^{3/2}}
    \label{eq:inverse_CT2}
\end{equation}
in which $f' = f'(R)$ is, as defined in Chapter \ref{chapter4}, the conformal transformation which leads from the JF to the EF ($\Tilde{g}_{\mu\nu} = f'g_{\mu\nu}$). We can then propose the following question: does the behavior of the Ricci scalar in one frame have to match the behavior of its counterpart in the other? More precisely, could a singularity be present in one frame but not in the other? To answer this question, we will analyze the case of a static and spherically symmetric source of curvature. We also assume an asymptotically flat spacetime. Based on these assumptions, we consider the radial component of the metric and $R$ to be of the following form:

\begin{align}
    &g_{rr} = 1 + \frac{1}{r^{\beta}},\quad \beta > 0 \label{eq:chap5_grr} \\
    &R = \frac{1}{r^m}. \quad m > 0 \label{eq:chap5_R(r)}
\end{align}

Notice that $R$ presents a physical singularity at $r = 0$. Calculating the covariant derivatives in (\ref{eq:inverse_CT2}) leads to
\begin{equation}
\Tilde{R} = \frac{R}{f'} - 6g^{rr}\frac{f''}{2 f'^{5/2}}\left[\left[\left(\frac{f'''}{f''} - \frac{1}{2f'^{1/2}}\right)\left(\frac{d R}{dr}\right)^2 + \frac{d^2 R}{d r^2}\right] - \Gamma_{rr}^r\frac{d R}{dr}\right],
\label{eq:inverse_CT3}
\end{equation}
and the Christofell symbol becomes
\begin{equation}
     \Gamma_{rr}^{r} = \frac{1}{2}g^{rr}\partial_{r} g_{rr}.
\end{equation}
\subsection{An Exponential $f(R)$}
For an exponential $f(R)$ we have
\begin{align}
    &f(R) = R + a e^{b R}
    \label{eq:exp_f(R)}, \quad (a>0,b>0) \\
    &f'(R) = 1+ a b e^{b R}, \\
    &f''(R) = a b^2 e^{b R}
    \label{eq:exp_f'}
\end{align}
where $a$ and $b$ are constants with dimensions of $R$ and $R^{-1}$ respectively.
Notice that $f'$, our conformal factor, diverges when $R \rightarrow \infty$ for $b > 0$, which corresponds to the limit $r \rightarrow 0$, where $R = 1/r^m$ is singular, and so we see that our CT is singular at the same point as $R$. The cosmological viability of some exponential models is discussed in \cite{amendola2007conditions} and is directly related to the study of the curves $K(r)$ in the $(r,K)$ plane, with $K \equiv R f''/f'$ and $r \equiv -R f'/f$. Recall that in Chapter \ref{chapter2} we discussed conditions on the derivatives of $f$ and concluded that $f'$ and $f''$ must be positive, which can be satisfied for (\ref{eq:exp_f(R)}) for $a > 0$.

For this $f(R)$, equation (\ref{eq:inverse_CT3}) yields
\begin{align}
    \tilde{R}(r) = &\frac{3 a b^2 e^{b r^{-m}} r^{\beta -2 m-2} (-2 a^3 b^4 e^{3 b r^{-m}}-6 a^2 b^3 e^{2 b r^{-m}}-6 a b^2 e^{b r^{-m}}+((a b e^{b r^{-m}}+1)^{5/2}-2 b)}{\left(r^{\beta }+1\right) \left(a b e^{b r^{-m}}+1\right)^{11/2}} \notag \\
    &-\frac{6 a b^2 \left(\beta +r^{\beta }+1\right) e^{b r^{-m}} r^{\beta -3 m-4}}{\left(r^{\beta }+1\right)^2 \left(a b e^{b r^{-m}}+1\right)^{5/2}} \notag \\
    &+\frac{r^{-m}}{a b e^{b r^{-m}}+1}.
    \label{eq:Re(r)_exp}
\end{align}
The interesting fact about (\ref{eq:Re(r)_exp}) is that, if $a>0$, $b>0$, $m>0$ and $\beta >0$, we have
\begin{equation}
    \lim_{r \rightarrow 0} \tilde{R}(r) = 0, 
    \label{eq:limit_exponential}
\end{equation}
showing us that the use of a conformal transformation that diverges exponentially as \(R \rightarrow \infty\) effectively removes the physical singularity of the JF Ricci scalar, ensuring that its EF counterpart, \(\tilde{R}\), remains well-defined at the origin. This is true if
\begin{equation}
\begin{aligned}
    a > 0, b>0, \\
    a < 0, b > 0.
\end{aligned}
\label{eq:exp_zero_conditions}
\end{equation}
The singularity is not removed if
\begin{equation}
\begin{aligned}
    a < 0, b < 0, \\
    a > 0, b < 0.
\end{aligned}
\label{eq:exp_singular_conditions}
\end{equation}
Notice that the case $(a>0,b>0)$ implies in $f'>0$ and $f''>0$, satisfying the conditions we saw in Chapter \ref{chapter2} for attractive gravity and stability with respect to perturbations of $R$, respectively.

The removal of the singularity happens because the exponential function of the $f(R)$ and of its derivatives diverges faster than powers of $1/r$, of $R(r)$ and its derivatives when $r\rightarrow 0$, which is equivalent to $R\rightarrow \infty$. Consider, for example, the term $R/f'$ on the right-hand side of (\ref{eq:inverse_CT3}),
\begin{equation}
    \frac{R}{f'} = \frac{R}{1+a b e^{b R}},
\end{equation}
which clearly goes to 0 when $R \rightarrow \infty$ for $b > 0$, with the same happening for the other terms inside the brackets.

This is similar to the case we explored in Chapter \ref{chapter3}, where we discussed how the coordinate singularity present in the Schwarzschild metric (\ref{eq:schwar_metric}), at $r = 2M$, can be removed by performing a coordinate transformation that is singular at that point, leading to the Kruskal-Szekeres coordinates, (\ref{eq:kruskal_metric}), in which the only singularity present is the physical one, at $r = 0$. The difference in this case, is that the singularity removed is that of the Ricci scalar, which is a physical one. 

Figure \ref{fig:ReXr_exp_comparision} shows the curve $\tilde{R} \times r$, where we can see how the EF Ricci scalar goes to zero as we approach what would be a singular point for $R$.
\begin{figure}[H]
    \centering
    \includegraphics[scale=0.41]{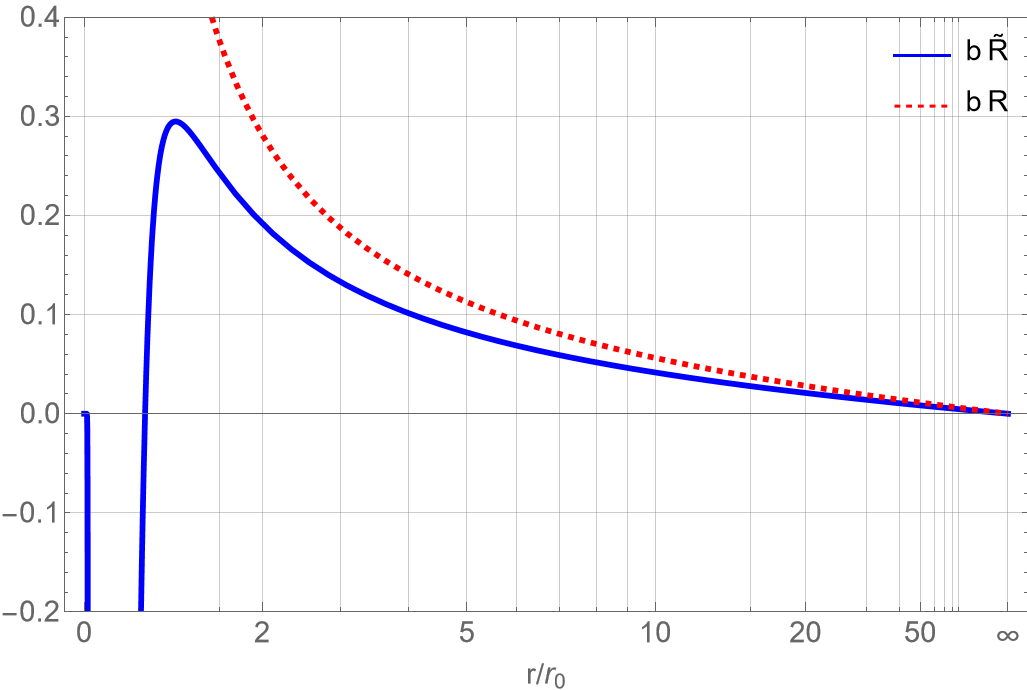}
    \caption{The EF Ricci scalar as a function of $r$ for $a = 1/3$, $b=1$, $\beta = 1$ and $m = 1$. In this frame, $\tilde{R}$ is finite at $r = 0$, in opposition to its JF counterpart, which is singular at that point. Equation (\ref{eq:exp_f(R)}) shows that $b$ has units of $1/R$, so $b R$ is dimensionless. The value $r_0$ corresponds to the value of the radial coordinate for which $\tilde{R}$ presents its global maximum.}
    \label{fig:ReXr_exp_comparision}
\end{figure}
Notice how, close to $r = 0$, there is a section of the curve that escapes the range of the vertical axis, figure \ref{fig:ReXr_exp2} displays the vicinity of $r = 0$, showing that $\tilde{R}$ also remains finite in that range.
\begin{figure}[H]
    \centering
    \includegraphics[scale=0.9]{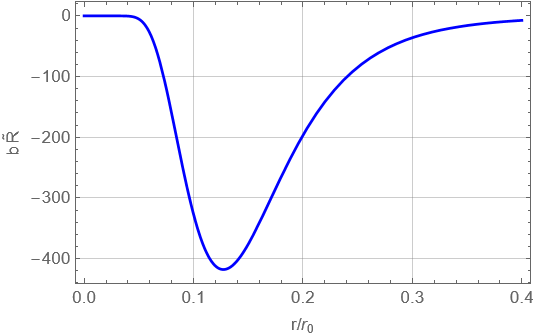}
    \caption{A zoom in the region close to $r=0$ of figure \ref{fig:ReXr_exp_comparision}, showing that $\tilde{R}$ also remains finite in this interval.}
    \label{fig:ReXr_exp2}
\end{figure}
We see here a case in which a physical singularity present in the JF is not manifested in the EF. In principle, this could be used as an argument against the physicality of the EF, but even if the singularity is removed, one could ask if the region close to $r = 0$ is ever probed by the dynamics of the system, perhaps the potential $V(\tilde{\phi})$ has some barrier that cannot be crossed, leaving the region which contains $r = 0$ inaccessible, a mechanism which would make up for the absence of singularity for $\tilde{R}(r=0)$. To answer this, we will analyze the behavior of $\tilde{\phi}$ and $V(\tilde{\phi})$ in a situation where $T^{(m)} = 0$, in which the effective potential, (\ref{eq:EF_potential_effective}), reduces to the original one from the EF action, (\ref{eq:EFaction}). This is satisfied in the case of a black hole and the outside of a star.\footnote{If we consider its interior, $T^{(m)} = 0$ is satisfied for relativistic matter, with the opposite case, for which $T^{(m)} \neq 0$, being that of matter at very low energies ($v/c \ll 1$ ). Such cases are governed by the chameleon effect \cite{khoury2004chameleon}. An interesting case, outside the scope of this dissertation, would be that of matter at intermediary speed, in which we would need an adequate radial profile for the density of matter, as well as equation of state, in the case of a star.}

Recalling that $\tilde{\phi} \equiv -\sqrt{3/2\kappa} \ln f'$ leaves us with the following expression for $\tilde{\phi}$:

\begin{equation}
    \tilde{\phi}(R) = -\sqrt{\frac{3}{2\kappa}}\ln(1 + a b e^{b R}).
    \label{eq:phi_til_exp}
\end{equation}
Notice that since $a$ and $b$ are positive, the argument of $\ln$ is always larger than $1$, causing the field to take up only negative values due to the minus sign.
Figure \ref{fig:phiE_exp} shows the curve $\tilde{\phi} \times R$, in which the aforementioned behavior is verified.
\begin{figure}
    \centering
    \includegraphics[scale=0.44]{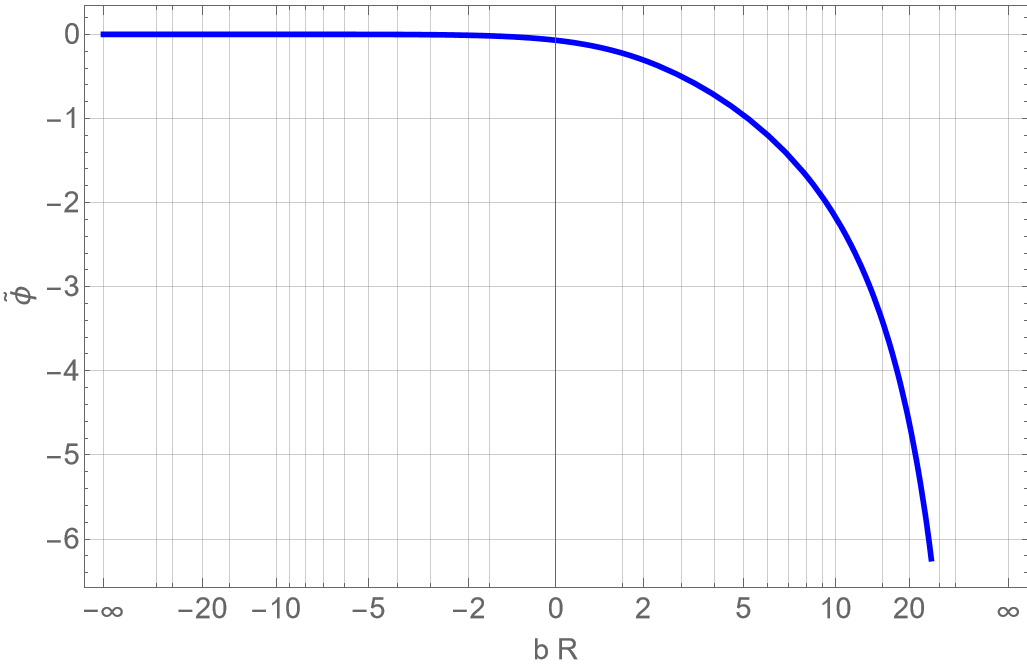}
    \caption{$\tilde{\phi}(R)$ for $a = 1/3$ and $b=1$. The conformal field takes only negative values and diverges to negative infinity}.
    \label{fig:phiE_exp}
\end{figure}
Recalling the definition of the EF potential, $V = U/2\kappa f'^2$, in terms of its JF conterpart, $U = R f' - f$, gives us
\begin{align}
    V(R) &= \frac{Rf'(R) - f(R)}{2\kappa f'(R)^2} \notag \\
    & = \frac{a e^{b R} (b R-1)}{16 \pi  \left(a b e^{b R}+1\right)^2},
    \label{eq:V_exp(R)}
\end{align}
and figure \ref{fig:VxR} shows the corresponding plot. The extreme points of the potential are given by the values of $R$ for which $dV/dR = 0$, corresponding to
\begin{equation}
   \frac{a b^2 e^{b R} \left(R-a e^{b R} (b R-2)\right)}{16 \pi  \left(a b e^{b R}+1\right)^3} = 0,
    \label{eq:exp_extremes}
\end{equation}
for which we found no analytical solution. 

We see that $V$ is always finite and presents a global maximum and minimum. Notice that the minimum being located at $R \neq 0$ is not consistent with the dependences assumed for $g_{rr}$ and $R$ in equations (\ref{eq:chap5_grr}) and (\ref{eq:chap5_R(r)}) where we assume that the universe is asymptotically flat.

Figure \ref{fig:VxPhi} shows a parametric plot of $V(R) \times \tilde{\phi}(R)$, in which we observe the same qualitative behavior as in the $V \times R$ plot, but in this case, the point $R = 0$ is located at the right side, corresponding to $\tilde{\phi} = 0$ and increases to the left side, towards more negative values of $\tilde{\phi}$.
\begin{figure}
    \centering
    \includegraphics[scale=0.44]{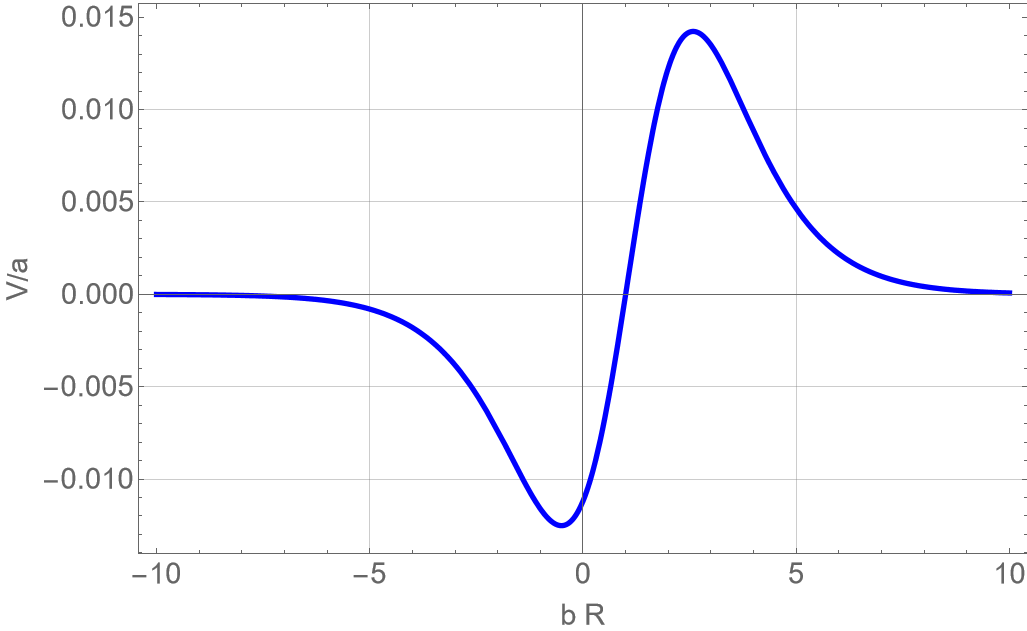}
    \caption{The behavior of the EF potential, V, as a function of R for $a = 1/3$ and $b=1$. The potential is always finite and presents a global maximum in the $R>0$ region and a global minimum in the $R<0$ region. Notice that $a$ has units of $R$, so $V/a$ is dimensionless. }
    \label{fig:VxR}
\end{figure}
\begin{figure}
    \centering
    \includegraphics[scale = 0.44]{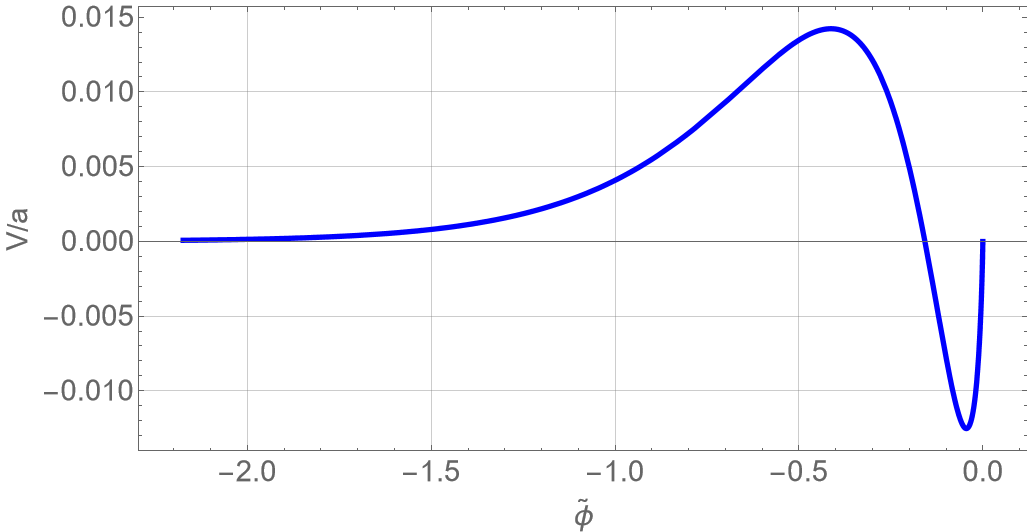}
    \caption{The parametric plot of $V$ as a function of $\tilde{\phi}$, for $a = 1/3$ and $b = 1$. Here we observe the same general behavior as in $V \times R$, but the figure is reflected. This means we have the point $R = 0$ located at the right side, corresponding to $\tilde{\phi} = 0$, increasing to the left side, towards more negative values of $\tilde{\phi}$.}
    \label{fig:VxPhi}
\end{figure}
In terms of the dynamics of $\tilde{\phi}$, figure \ref{fig:VxPhi} shows that depending on its energy, the field could cross the potential barrier and the system would be free to evolve towards larger values of $R$ (more negative values of $\tilde{\phi})$, but without ever reaching a singularity in the EF. If the energy is not large enough, $\tilde{\phi}$ will oscillate around the global minimum, and the system would never reach values of $R$ that lie beyond the barrier. In terms of the JF, a possible conjecture is that this could be a mechanism that could prevent the creation of singularities in the collapse of stars, for example. If the initial conditions at the star are such that the conformal field has enough energy to cross the barrier, the system would be allowed to evolve to larger and larger values of $R$, eventually generating a black hole and a singularity. If not, then the the creation of a singularity would be forbidden, because $\tilde{\phi}$ would never cross the potential barrier. The study of this conjecture, however, is out of the scope of this dissertation and is left as a possibility for future work.
\subsection{A Starobinsky-like $f(R)$}
For a Starobinsky-like $f(R)$ \cite{starobinsky1980new} we have 
\begin{align}
    &f(R) = R + \alpha R^n, \label{eq:starobinsky} \\
    &f'(R) = 1 + \alpha n R^{n-1}, \\
    &f''(R) = \alpha n(n-1)R^{n-2},
\end{align}
where we see that, as in the exponential $f(R)$ case, the conformal factor $f'$ diverges as $R \rightarrow \infty$ if $n > 1$. The cosmological viability of Starobinsky-like models is discussed in \cite{amendola2007conditions}. Expressing $\tilde{R}$ as a function of $r$ leaves us with
\begin{align}
    \tilde{R}(r) = -&\frac{6 \alpha  (n-1) n \left(\beta +(n-2) r^{\beta +2 m+2}+(n-2) r^{2 m+2}+r^{\beta }+1\right) r^{\beta -m (n+1)-4}}{\left(r^{\beta }+1\right)^2 \left(\alpha  n r^{m-m n}+1\right)^{5/2}} \notag \\ 
    &+\frac{1}{\alpha  n r^{-m (n-2)}+r^m} \notag \\
    &+\frac{3 \alpha  (n-1) n r^{\beta +2 m n-2}}{\left(r^{\beta }+1\right) \left(r^{m n}+\alpha  n r^m\right)^3},
    \label{eq:R(r)_starobinsky}
\end{align}
and in the limit $r \rightarrow 0$ we have
\begin{equation}
    \displaystyle \lim_{r \rightarrow 0} \tilde{R}(r) = 0,
\end{equation}
if the following conditions are satisfied simultaneously
\begin{align}
\label{eq:staro_zero_conditions}
\left\{\begin{array}{l}
\alpha>0, \\
n>2, \\
\beta >1, \\
\beta + 2mn >2, \\
7m + 8 < 2\beta + 3mn, \\
\beta + m < mn, \\
3m + 8 < 2\beta + mn, \\
\beta < 2m + 2, \\
4\beta < mn + m + 8.
\end{array}\right. 
\end{align}
Notice that $(\alpha > 0, n > 2)$ lead to $f'>0$ and $n>2$, satisfying the conditions seen in Chapter \ref{chapter2} for the derivatives of $f$.
If the previous conditions are violated, $\tilde{R}$ remains singular whenever a set of the following group of conditions is satisfied
\begin{align}
 &\left\{\begin{array}{l}
\alpha>0, \label{eq:staro_singular_conditions1}\\
n < 1, \\
\beta < 4 + m + mn, \\
2mn + \beta > 2, \\
\beta < 2 + 2m, \\ 
m + \beta < 2 + mn, \\
m < mn + \beta,
\end{array}\right. \\
&\left\{\begin{array}{l}
\alpha > 0, \\
n < 0, \\
\beta < 4 + m + mn, \\
2mn + \beta < 2, \\
m < mn + \beta, \\
m + \beta < 2 + mn, \\
\beta < 2 + 2m,
\end{array}\right. \\
 &\left\{\begin{array}{l}
\alpha<0, \\
n<0, \\
\beta < 4 + m + mn, \\
2mn + \beta < 2, \\
m < mn + \beta, \\
m + \beta < 2 + mn, \\
\beta < 2 + 2m,
\end{array}\right.
\end{align}
\begin{align}
 &\left\{\begin{array}{l}
\alpha<0, \\
n < 1, \\
\beta < 4 + m + mn, \\
2mn + \beta > 2, \\
\beta < 2 + 2m, \\
m + \beta < 2 + mn, \\
m < mn + \beta,
\end{array}\right. & \\
 &\left\{\begin{array}{l}
\alpha >0, \\
1<n<2, \\
2mn + \beta > 2, \\
3mn + 2\beta < 8 + 7m, \\
mn < m + \beta, \\
mn + \beta < 2 + m, \\
\beta < 2 + 2m. \label{eq:staro_singular_conditions2}
\end{array}\right. &
\end{align}
Figure \ref{fig:f(R)_region} shows the corresponding region plots, with the blue region corresponding to a non-singular  $\tilde{R}$, and the red region being the opposite case. Notice how the the singularity is removed only if $n > 2$, meaning that only in this case the conformal factor $f'$ is able to diverge faster than the powers of $1/r$ of $R(r)$ and its derivatives. Figure \ref{fig:ReXr_staro_comparision} exhibits the behavior of $\tilde{R}$ and $R$ as a function of $r$, illustrating how the singularity is absent in $\tilde{R}$. A zoom in the region close to $r = 0$ is displayed in figure \ref{fig:ReXr_staro_zoom}, showing that $R$ is always finite.
\begin{figure}[H]
  \centering
  \begin{subfigure}[b]{0.45\textwidth}
    \includegraphics[width=\textwidth]{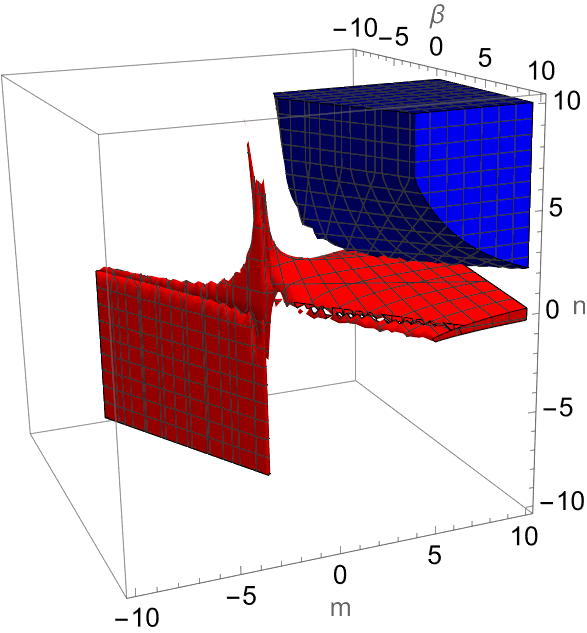}
  \end{subfigure}
  \hfill
  \begin{subfigure}[b]{0.45\textwidth}
    \includegraphics[width=\textwidth]{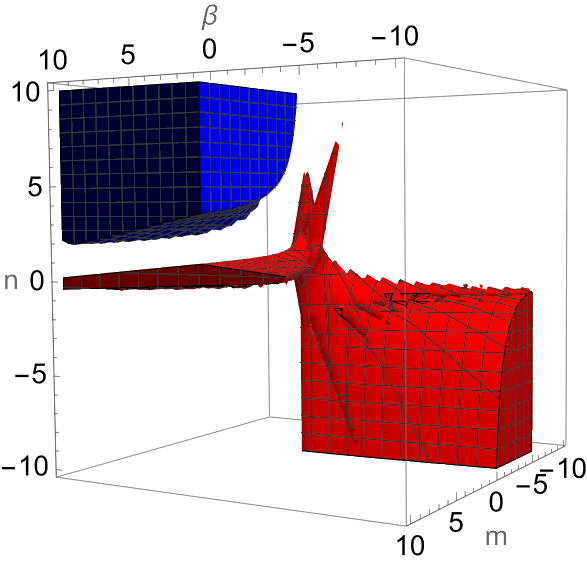}
  \end{subfigure}
  \caption{Two different views of the region plot corresponding to conditions (\ref{eq:staro_zero_conditions}) and (\ref{eq:staro_singular_conditions1})-(\ref{eq:staro_singular_conditions2}). In the blue region $\tilde{R}$ is not singular, while in the red region it remains so. Notice that only for $n > 2$ the singularity can be removed.}
  \label{fig:f(R)_region}
\end{figure}
\begin{figure}[H]
    \centering
    \includegraphics[scale=0.44]{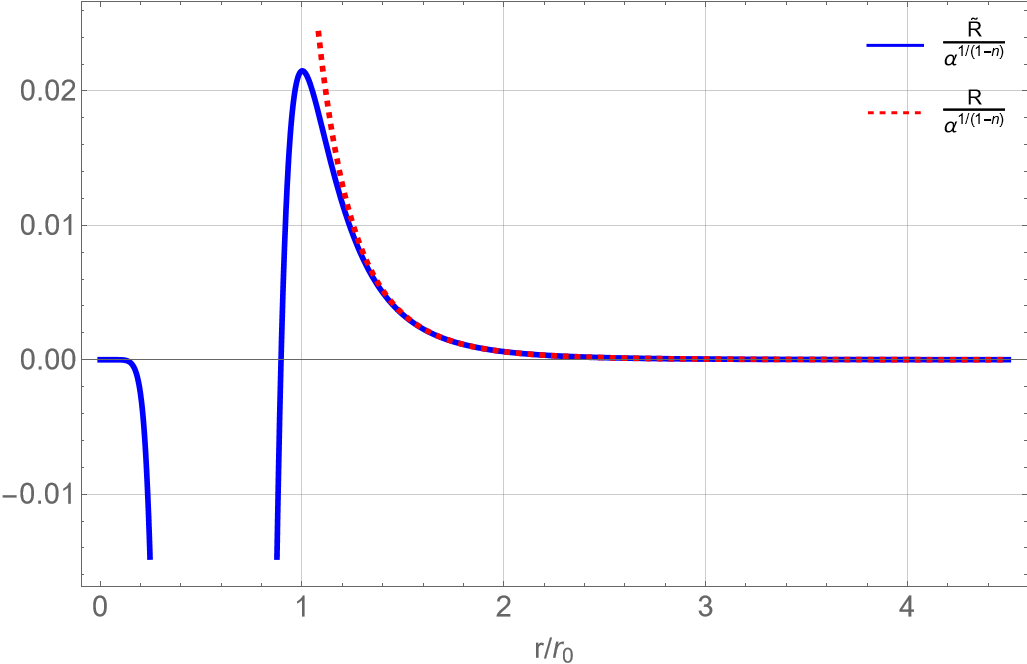}
    \caption{Here we see the behavior of $\tilde{R}$ and $R$ as a function of $r$ for $m = 6$, $\beta = 6$, $n = 3$ and $\alpha = 1$, illustrating how the singularity is absent in $\tilde{R}$. Notice that, as in the exponential $f(R)$ case, $\tilde{R}$ is always finite in the interval $[0,3]$, not covered by vertical axis. Equation (\ref{eq:starobinsky}) shows that $\alpha^{1/(1-n)}$ has units of $R$, so $R/\alpha^{1/(1-n)}$ is dimensionless. The value $r_0$ corresponds to the value of the radial coordinate for which $\tilde{R}$ presents its global maximum.}
    \label{fig:ReXr_staro_comparision}
\end{figure}
\begin{figure}[H]
    \centering
    \includegraphics[scale=0.44]{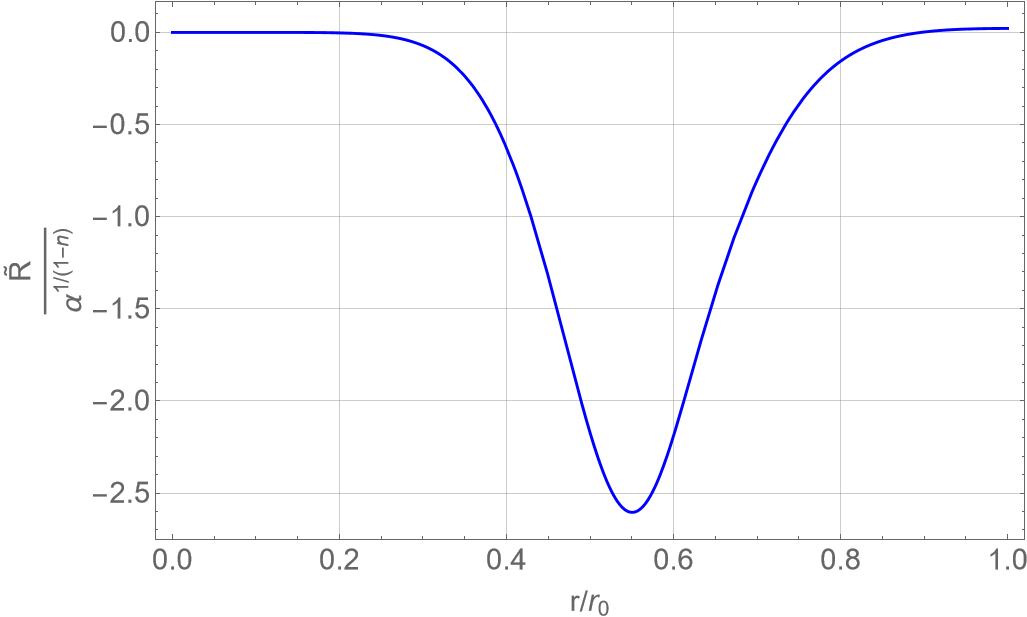}
    \caption{Here we see a zoom in the region close to $r = 0$, emphasizing that $\tilde{R}$ is always finite.}
    \label{fig:ReXr_staro_zoom}
\end{figure}
As done in the case of an exponential $f(R)$ we will now look at the behavior of the conformal field and the potential. The expression for the field is
\begin{equation}
    \tilde{\phi}(R) = -\sqrt{\frac{3}{2\kappa }} \ln \left(\alpha  n R^{n-1}+1\right).
    \label{eq:phi_staro}
\end{equation}
Notice that for an odd value of $n$, the argument of the $\ln$ function is always positive if $\alpha > 0$ and $n > 0$, meaning $\tilde{\phi}$ will always be negative. For an even $n$, there will be a critical value of $R$, after which the argument becomes negative, yielding a complex field. This value is given by 
\begin{equation}
    R^* = -\left(\frac{1}{\alpha n}\right)^{1/(n-1)}.
    \label{eq:critical}
\end{equation}
In figure \ref{fig:phiE_staro} we see the plot $\tilde{\phi} \times R$ for $n = 3$ and $\alpha = 1$. It exhibits a global maximum at $R = 0$, where $\tilde{\phi}(0) = 0$. The field diverges to negative infinity when $R \rightarrow \pm \infty$.
\begin{figure}[H]
    \centering
    \includegraphics[scale=0.44]{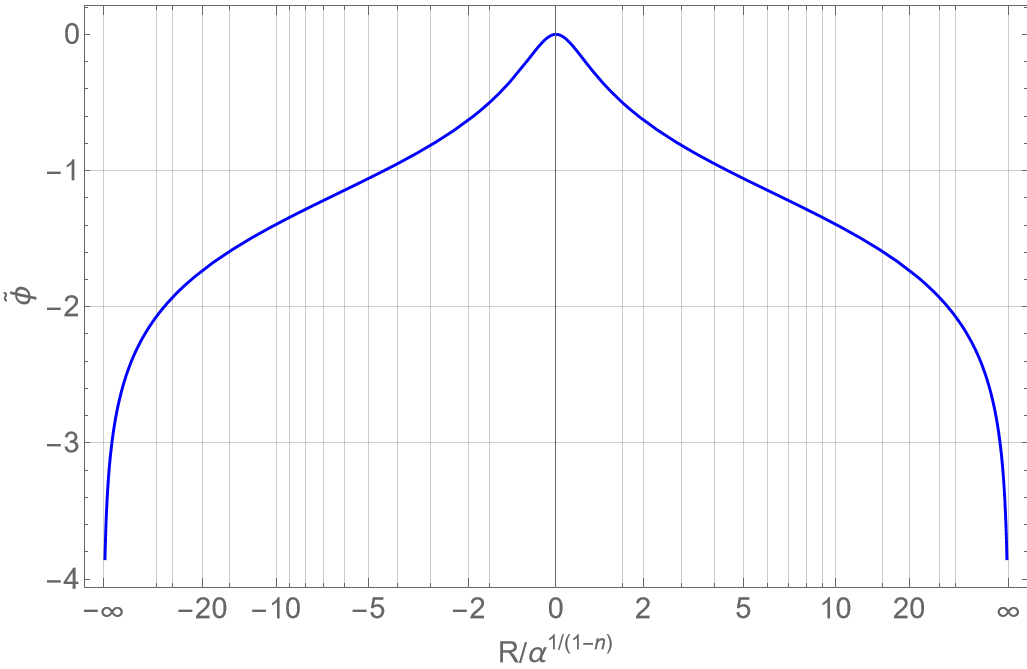}
    \caption{$\tilde{\phi} \times R$ for $n = 3$ and $\alpha = 1$. Here we see that the field has a global maximum at $R = 0$, where it is equal to zero. We see also that $\tilde{\phi}$ diverges to negative infinity when $R \rightarrow \pm \infty$. }
    \label{fig:phiE_staro}
\end{figure}
The potential takes up the form
\begin{align}
    V(R) = \frac{\alpha  (n-1) R^{n+2}}{2 \kappa  \left(\alpha  n R^n+R\right)^2},
    \label{eq:V_staro(R)}
\end{align}
with its extreme points being the solution for $dV/dR = 0$, corresponding to
\begin{equation}
  \frac{\alpha  (n-1) n R^{n+1} \left(\alpha  (n-2) R^n-R\right)}{16 \pi  \left(\alpha  n R^n+R\right)^3} = 0.
    \label{eq:staro_extremes}
\end{equation}
In figure \ref{fig:VxR_staro} we see the potential curve, which exhibits a global maximum in the $R>0$ region and a global minimum in the $R<0$ region. Figure \ref{fig:VxPhi_staro} shows the parametric plot $V(R) \times \tilde{\phi}(R)$ and we see that it has two branches. This happens because for an odd $n$, the power of $R$ in equation (\ref{eq:phi_staro}) will be an even one, causing symmetric values of $R$ to yield the same value of $\tilde{\phi}$. The upper branch displays a potential barrier and is associated to $R>0$, while the lower branch has a potential well and is associated to $R<0$. Here the question of which branch is the physical one arises.

The issue of multivalued potentials is discussed in \cite{frolov2008singularity} and \cite{kobayashi2008relativistic}, but the work is restricted to the JF and the point of ramification corresponds to a singularity of the Ricci scalar, making it impossible to cross from one branch to another. This is not the case here, since it can be seen in figure \ref{fig:VxPhi_staro} that the point of ramification occurs at $\tilde{\phi} = 0$, which corresponds to $R = 0$, as shown in figure \ref{fig:phiE_staro}. It was also verified that if $n$ is not an integer number, then the issue of two branches disappears and we are left with the upper branch. This happens because the power of $R$ in equation (\ref{eq:phi_staro}) yields a complex number when $R < 0$. This is shown in figure \ref{fig:VxPhi_staro_decimal}, which displays $V \times \tilde{\phi}$ for $n = 3.1$, and we see that the lower branch is absent. 

Choosing the correct branch is a matter of analysing which one is associated to an acceptable cosmological solution, this however, is outside the scope of this dissertation and is left a possibility for future development of this work. The fact is that the arguments used in the case of an exponential $f(R)$ are still valid here for whichever branch we choose, with the sole difference being that this potential has no local minimum close to $R = 0$ around which $\tilde{\phi}$ can oscillate. Still, depending on the field's energy, it could cross either the potential barrier (upper branch) or the potential well (lower branch), and the system would be able to reach larger and larger values of $R$ without ever running into a singularity in the EF. Again, in the perspective of the JF, the conjecture proposed in the previous case remains valid, and this could act as a mechanism through which the collapse of a star would result in a singularity or not.
\begin{figure}[H]
    \centering
    \includegraphics[scale=0.5]{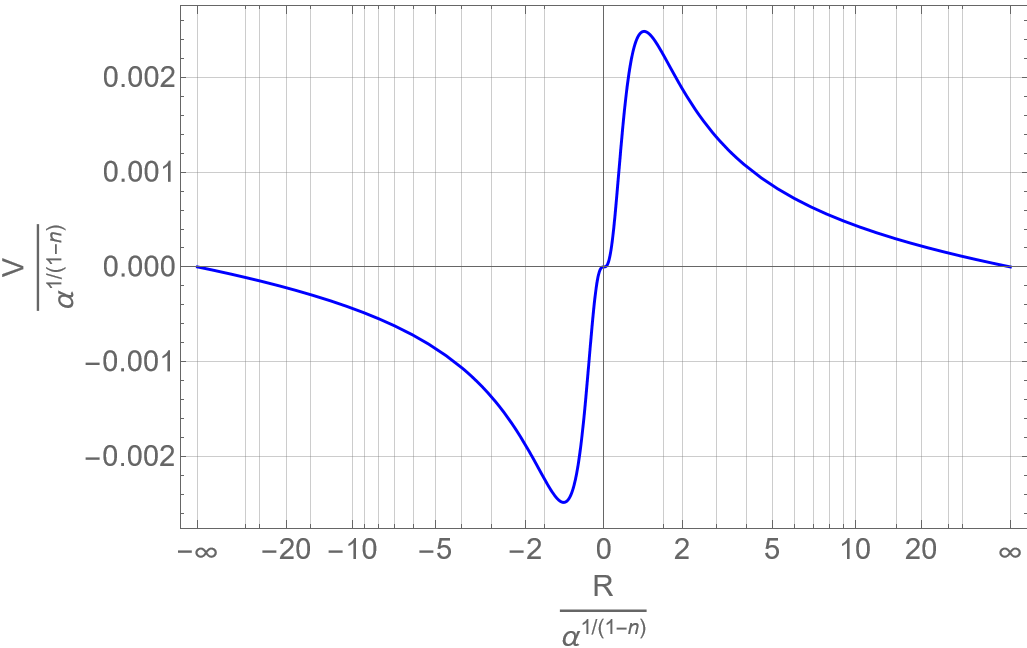}
    \caption{The behavior of the EF potential, $V$, as a function of $R$ for $n = 3$ and $\alpha = 1$. It has a global maximum in the region associated to $R>0$ and a global minimum in the $R<0$ region.}
    \label{fig:VxR_staro}
\end{figure}
\begin{figure}[H]
    \centering
    \includegraphics[scale=0.5]{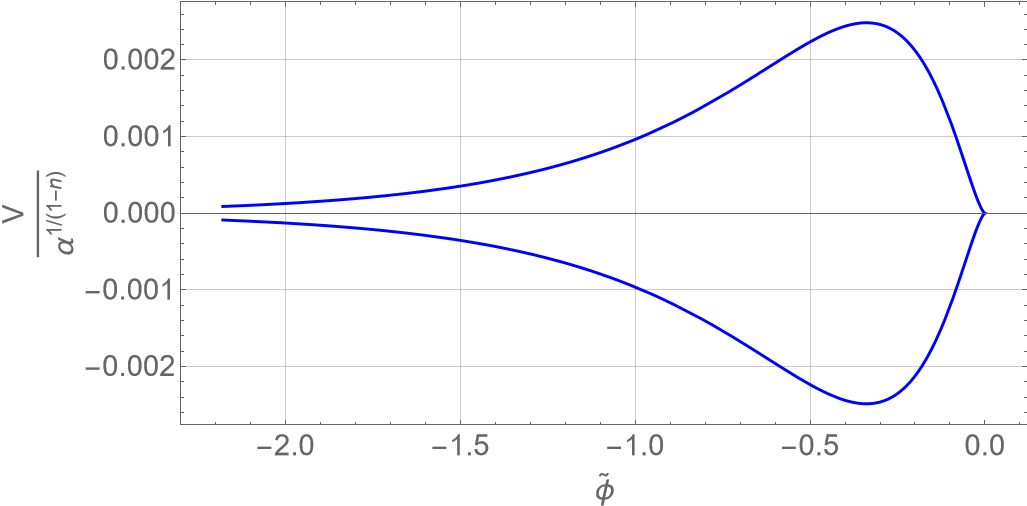}
    \caption{The behavior of the EF potential, $V$, as a function of $\tilde{\phi}$ for $n = 3$ and $\alpha = 1$. We see that it has two branches. The upper branch is associated to positive values of $R$, while the lower branch, to negative values.The issue of branches arises for every odd value of $n$.}
    \label{fig:VxPhi_staro}
\end{figure}
\begin{figure}[H]
    \centering
    \includegraphics[scale=0.44]{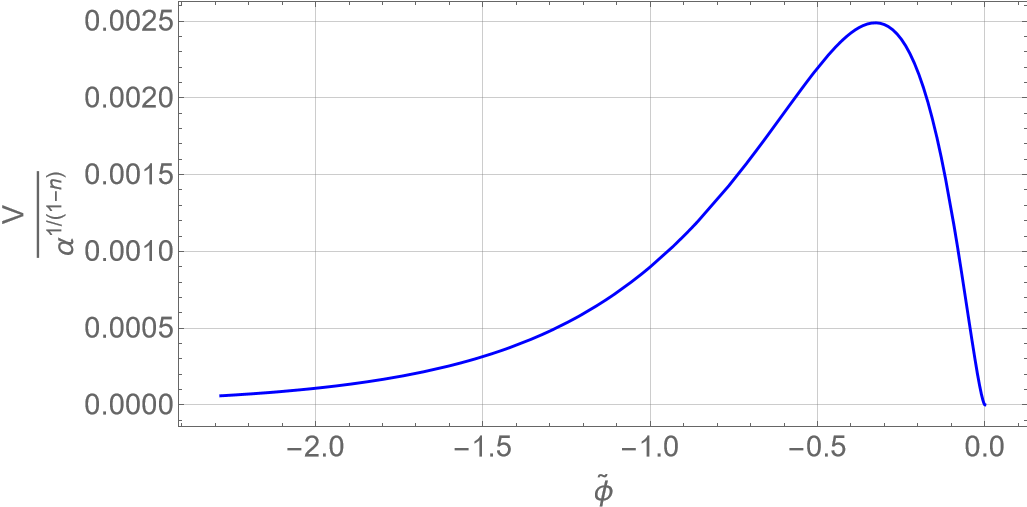}
    \caption{The behavior of the EF potential, $V$, as a function of $\tilde{\phi}$ for $n = 3.1$ and $\alpha = 1$. $n$ being a decimal number removes the lower branch. This was observed for a variety of non-integer values of $n$.}
    \label{fig:VxPhi_staro_decimal}
\end{figure}
Another interesting case is obtained when we consider an even value for $n$. Here, equation (\ref{eq:phi_staro}) predicts positive values for the field when $R$ is negative, opposing the previous cases in which $\tilde{\phi}$ assumed only negative values. We established in equation (\ref{eq:staro_zero_conditions}) that the minimum condition for $n > 2$ is needed for the singularity to be removed in the EF. Let us then consider the case $n = 4$. Figure \ref{fig:phiE_staro_even} shows the corresponding plot for $\tilde{\phi}$ as a function $R$. Notice that in this case, for $R < 0$, the field has positive values and diverges to positive infinity when it reaches a critical value, $R^*$, given by equation (\ref{eq:critical}), after which it becomes complex-valued.

Figure \ref{fig:VxR_staro_even} displays the corresponding potential $V$ as a function of $R$. Notice that in opposition to the cases with $n = 3$ and $n = 3.1$, here we observe a global minimum at $R = 0$, corresponding to a Minkowski-like universe, in accordance to our assumption that the metric is asymptotically flat in equation (\ref{eq:chap5_grr}). Notice that a singularity in the potential can be seen in the $R<0$ region in figure \ref{fig:VxR_staro_even}, leaving part of that region inaccessible. This divergence happens at the same critical value, $R^*$, after which $\tilde{\phi}$ becomes complex-valued causing the potential be no longer physical after this value.

Figure \ref{fig:VxPhi_staro_even} exhibits $V$ as a function of $\tilde{\phi}$ and the same qualitative behavior as in figure \ref{fig:VxR_staro_even} is observed. This plot covers the accessible region associated to $R > R^*$.  The minimum is located at $R = 0$ and the potential goes to 0 as $R$ diverges to positive infinity. The conjecture proposed in the previous cases remains valid. Notice that the existence of a minimum at $R = 0$ is only possible because the field $\tilde\phi$ is not restricted to negative values when $n$ is an even number. This is the reason why in every other case a minimum at $R = 0$ is absent, since $\tilde{\phi}$ there is always negative.\footnote{This is compatible with the requirement that $f''>0$ so that perturbations around $R = 0$ will not generate an unbounded growth of curvature, as discussed in Chapter \ref{chapter2}. 
}
\begin{figure}[]
    \centering
    \includegraphics[scale=0.44]{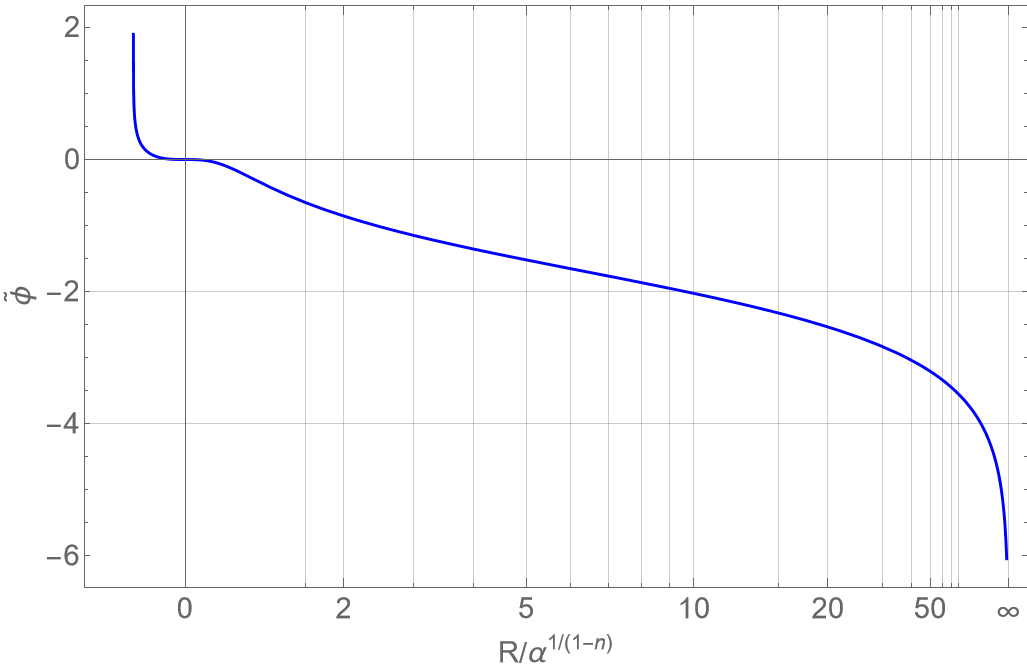}
    \caption{$\tilde{\phi} \times R$ for $n = 4$ and $\alpha = 1$. For $R < 0$, the field has positive values and diverges $+\infty$ when it reaches the critical value $R^* = -0.62$, after which it becomes complex-valued. For $R>0$ it is negative and diverges to $-\infty$.}
    \label{fig:phiE_staro_even}
\end{figure}
\begin{figure}[]
    \centering
    \includegraphics[scale=0.44]{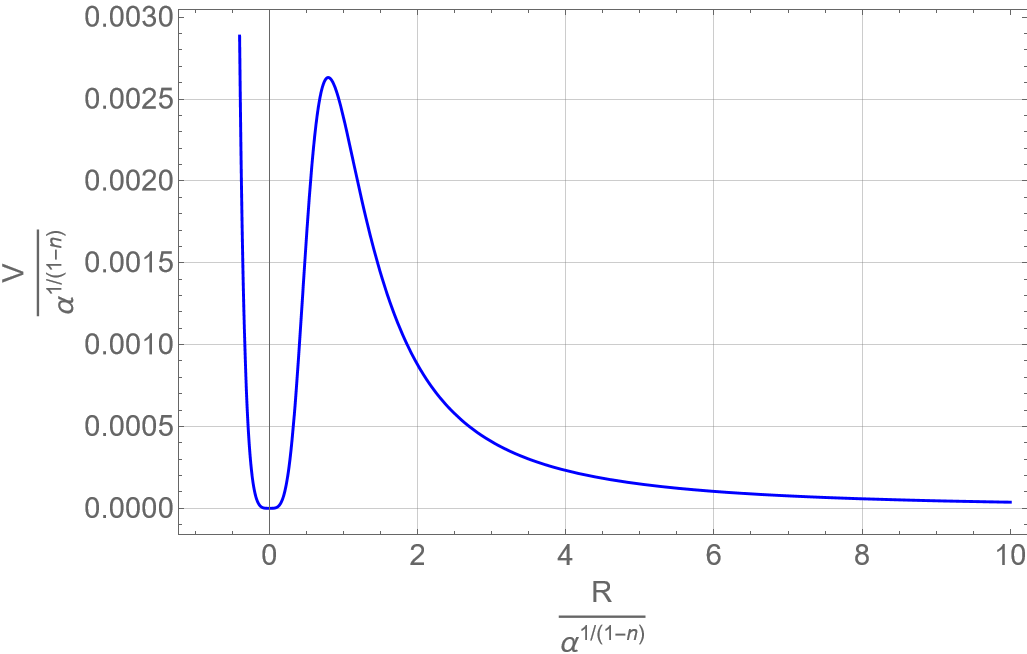}
    \caption{The behavior of the EF potential, $V$, as a function of $R$ for $n = 4$ and $\alpha = 1$. A singularity is present at $R^* = -0.62$, where equation (\ref{eq:V_staro(R)}) diverges, leaving the region for $R$ more negative than that inaccessible. This is the same value of $R$ after which $\tilde{\phi}$ becomes complex-valued. }
    \label{fig:VxR_staro_even}
\end{figure}
\begin{figure}[]
    \centering
    \includegraphics[scale=0.44]{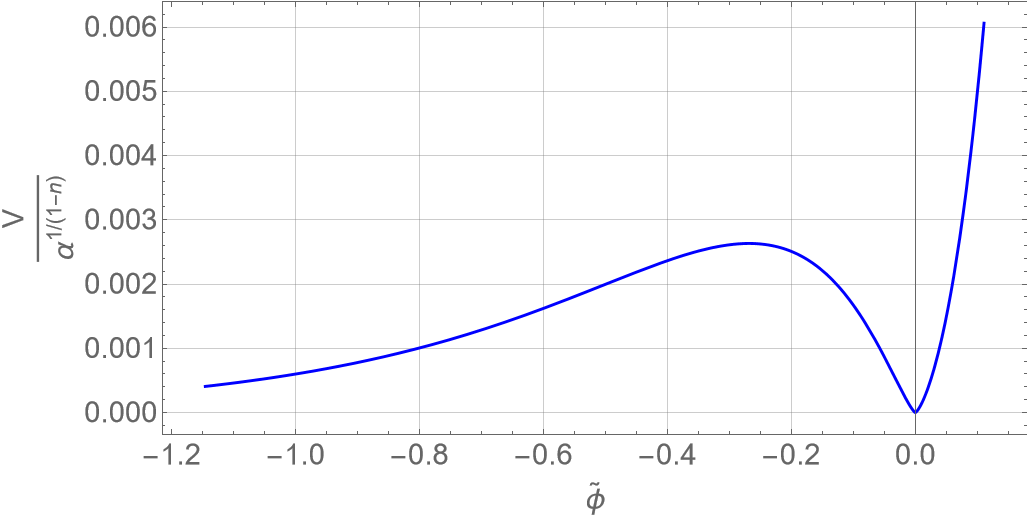}
    \caption{The behavior of the EF potential, $V$, as a function of $\tilde{\phi}$ for $n = 4$ and $\alpha = 1$ for the accessible region associated to $R > R^* = -0.62$. For every even value of $n$, there will be a singularity that leaves part of the $R<0$ region inaccessible and the minimum will be located at $R = 0$. }
    \label{fig:VxPhi_staro_even}
\end{figure}

We have seen how a CT that is singular at the same point as the JF Ricci Scalar can remove the singularity of its EF counterpart. The presence of a physical singularity in one frame and its absence in the other could be used as an argument against the physical equivalence of the frames, but this is far from a definitive proof on this matter. If we choose to interpret the results obtained in the EF through the perspective of the JF, it is then possible to conjecture a physical situation in which the creation of singularities in the JF can be prevented depending on the dynamics of the EF conformal field. If $\tilde{\phi}$ has enough energy to cross the potential barrier, the dynamics allow it to assume values that correspond to arbitrarily large values of the JF Ricci scalar, $R$. This is possible because of the absence of a singularity in the EF Ricci scalar, $\tilde{R}$. If $\tilde{\phi}$ does not have enough energy to cross the potential barrier, then it may oscillate around the global minimum, if there is one.

We have seen that in the case of a Starobinsky-like $f(R)$ with odd or non-integer values of $n$, the potential $V$ presents no global minimum around which $\tilde{\phi}$ can oscillate. This is problematic, since it indicates the absence of stable cosmological solutions \cite{ivanov2012stable}\cite{sawicki2007stability}. For for the $n = 4$ and exponential $f(R)$ cases we observed a minimum located at $R = 0$ in the former and at $R \neq 0$ in the latter. The existence of a minimum at null curvature is consistent with our initial assumption of an asymptotically flat spacetime, while it being located at $R \neq 0$ is not. 

Further investigation is needed with respect to the cosmological viability of the $f(R)$ models utilized, and also with respect to the results obtained in the EF and its interpretation in terms of the JF. The possibility to eliminate a physical singularity present in a JF scalar quantity of its EF counterpart paves the way to the discussion on the physical equivalence of the frames and gives rise to an interesting conjecture on how the dynamics of the conformal field could be used as a mechanism to better understand the creation of physical singularities in the JF.

Even if the physical equivalence between the frames is not at all an obvious matter, what we can say is that each frame has its own advantages. The familiarity of the EF field equations (which are those of GR with the presence of a field $\tilde{\phi}$) and the fact that physical singularities present in the JF do not necessarily manifest themselves on the EF can be used as an argument for one to make calculations in the EF. However, since the conservation of $T_{\mu\nu}^{(m)}$ is only verified in the JF, at the end of the day one has to return to this frame in order to analyze the physical consequences of results obtained in the EF. One, however, must be aware that it may be done at the cost of running into a physical singularity which was not present in the EF. This choice of the initial  frame (in particular, picking the more familiar and intuitive one) is analogous to the situation we described in Chapter \ref{chapter3}. There, we stated that one is free to begin the calculations in the Kruskal coordinates (where only the physical singularity is present). We also pointed out that the reason we usually work with the Schwarzschild line element is that the spherical coordinates are familiar and more intuitive, even if we have to pay the price of having a coordinate singularity at $r = 2M$.

\end{chapter}

\begin{chapter}{Conclusions}
\label{conclusions}
In this dissertation we discussed the intricacies of conformal transformations in the context of $f(R)$ gravitation and the equivalence two frames in which the theory can be cast. We began by introducing the fundamental field equations of General Relativity and of $f(R)$ theories. It was shown that the latter presents an extra scalar degree of freedom, with the Ricci scalar, $R$, being related to the trace of the energy-momentum tensor, $T$ through a differential equation, in opposition to general relativity, where this relation is algebraic. Conditions on the first and second derivatives of the $f(R)$ functions which must be satisfied in order for the theory to produce stable cosmological solutions were discussed.

We then moved on to a discussion on coordinate transformations in which we stated the difference between physical and coordinate singularities, showing how the latter can be removed from the line element by means of a convenient coordinate transformation. Specifically, we discussed the presence of a coordinate singularity in the Schwarzschild metric and how it leads one to believe that a test particle never crosses the event horizon of the black hole, which is a null surface located exactly at the point where the coordinate singularity is present. A convenient coordinate transformation was utilized, casting the line element in the Kruskal-Szekeres coordinates, in which the coordinate singularity is absent and only the physical one remains. This makes clear that a radially infalling particle will not only cross the event horizon at finite proper time, but that once it does, it will inevitably reach the singularity.

We then proceeded to discuss conformal transformations in the context of $f(R)$ theories. We explored their mathematical properties and the associated physical consequences, such as the preservation of the causal structure of spacetime due to the fact that CTs preserve angles. A classic example of the usefulness of CTs in physics was examined, in which it was shown how the preservation of angles allows one to cast difficult problems in more convenient geometrical formulations where the solution is easier to obtain. After this the application of CTs to $f(R)$ gravitation was discussed.

It was shown how the theory can be expressed in two mathematically equivalent frameworks, the JF and EF, connected by a CT of the metric. The physical equivalence of the frames was investigated and it was shown how the conservation the energy-momentum tensor, $T_{\mu\nu}^{(m)}$ and the geodesic equation of a free-falling particle both being modified does not imply in a physical inequivalence. That happens because a natural consequence of the coupling of the conformal field to the matter action causes the units of the EF to become a function of the spacetime point, being referred to as running units. The running of units allows physical laws to be conformally invariant, as well as experiments which measure ratios between some observable and its unit measure to yield results that agree in both frames.

Finally, it was shown that the physical singularity of the JF Ricci scalar can be removed from its EF counterpart by means of a CT that is singular in the same point as $R$ but diverges faster. This was shown for two $f(R)$ models. In the first one we considered the function $f(R) = R + a \exp(b R)$, an exponential model. In the second we considered $f(R) = R + \alpha R^n$, a Starobinsky-like model. We assumed an asymptotically flat metric and $R = 1/r^m$ ($m>0$) as the dependence of $R$ on the radial coordinate $r$.

For the exponential case it was verified that the EF Ricci scalar, $\Tilde{R}$, is not singular if $(a>0,b>0)$ or $(a<0, b>0)$. A parametric plot of the EF potential, $V$, as a function of the conformal field, $\tilde{\phi}$, shows that the potential has a minimum at a negative value of $R$ and a potential barrier in the $R > 0$ region, after which the potential goes to zero as $R \rightarrow \infty$. This plot shows that the dynamics of the field may allow it to cross the potential barrier and assume values that correspond to arbitrarily large values of $R$, if enough energy is available. If this does not happen, the field may oscillate around the minimum. Notice that the minimum of the potential being located a negative value of $R$ is not consistent with our assumption of an asymptotically flat spacetime.

For the Starobinsky-like $f(R)$ we verified that $\tilde{R}$ is not singular if $n > 2$, among other conditions involving inequalities between $\alpha$ and parameters present in the metric and in $R$ (see eq. (\ref{eq:staro_zero_conditions})). We analyzed the cases $n = 3$, $n = 3.1$ and $n = 4$. For the $n = 3$ case, the plot $V \times \tilde{\phi}$ showed that the potential has two branches. One with a global maximum associated to positive $R$ (potential barrier) and one with a global minimum associated with negative $R$ (potential well). A local minimum around which $\tilde{\phi}$ can oscillate if it cannot cross the barrier or the well (depending on which is the physical branch) is absent. This is problematic since it indicates the absence of stable cosmological solutions. The matter of which branch is physical is left as a possibility for future investigations. This behavior happens for any odd value of $n$. For $n = 3.1$ we verified that the only difference is that the branch corresponding to $R < 0$ is removed, leaving only the potential barrier associated with $R > 0$. Still, no local minimum around which $\tilde{\phi}$ could oscillate is present. This happens for every non-integer value of $n$.

For $n = 4$, the plot $V \times \tilde{\phi}$ exhibits the usual potential barrier associated to $R > 0$ and a global minimum located at $R = 0$, which is consistent with our assumption of an asymptotically flat spacetime. This happens for every even value of $n$. The possibility of a physical singularity being present in one frame but not in the other could be used as an argument against their physical equivalence, however, assuming the equivalence to be true allows us to interpret the results obtained in the EF through the perspective of the JF and propose the following conjecture: the absence of the Ricci scalar singularity in the EF, along with the fact that the dynamics of $\tilde{\phi}$ allow it to cross the potential barrier and take up values corresponding to arbitrarily large values of $R$ or oscillate around a minimum depending on its energy, can be seen as a mechanism associated to the creation of singularities in the collapse of stars. If the initial conditions of the star are such that $\tilde{\phi}$ has enough energy to cross the barrier, it would then be free to reach arbitrarily large values of $R$, which in the JF, means a black hole and a singularity would be created. If there is not enough energy, then the field would not cross the potential barrier and the collapse of the star would not create a singularity. The detailed investigation of the proposed conjecture is left as a possibility for future work. We also intend to investigate the case where $T^{(m)} = 0$, for which we would need a radial profile for the density of a star, as well as an equation of state for the relation between pressure and matter density.

A similar analysis could be conducted using other scalar quantities, like the other algebraic invariants of the Riemann tensor \cite{harvey1990algebraic}. This is also left a possibility of future work.

\end{chapter}

\addcontentsline{toc}{chapter}{Bibliography}
\printbibliography


\end{document}